# Likelihood-free inference via classification

**Michael U. Gutmann · Ritabrata Dutta · Samuel Kaski · Jukka Corander**

**Abstract** Increasingly complex generative models are being used across disciplines as they allow for realistic characterization of data, but a common difficulty with them is the prohibitively large computational cost to evaluate the likelihood function and thus to perform likelihood-based statistical inference. A likelihood-free inference framework has emerged where the parameters are identified by finding values that yield simulated data resembling the observed data. While widely applicable, a major difficulty in this framework is how to measure the discrepancy between the simulated and observed data. Transforming the original problem into a problem of classifying the data into simulated versus observed, we find that classification accuracy can be used to assess the discrepancy. The complete arsenal of classification methods becomes thereby available for inference of intractable generative models. We validate our approach using theory and simulations for both point estimation and Bayesian inference, and demonstrate its use on real data by inferring an individual-based epidemiological model for bacterial infections in child care centers.

## 1 Introduction

The likelihood function plays a central role in statistical inference by quantifying to which extent some values of the model parameters are consistent with the observed data. For complex models, however, evaluating the likelihood function can be computationally very costly, which often prevents its use in practice. This paper is about statistical inference for generative models whose likelihood-function cannot be computed in a reasonable time.[1]

A generative model is here defined as a parametrized probabilistic mechanism which specifies how the data are generated. It is usually implemented as a computer program that takes a state of the random number generator and some values of the model parameters $\boldsymbol{\theta}$ as input and that returns simulated data $\mathbf{Y}_{\boldsymbol{\theta}}$ as output. The mapping from the parameters $\boldsymbol{\theta}$ to simulated data $\mathbf{Y}_{\boldsymbol{\theta}}$ is stochastic and running the computer program for different states of the random number generator corresponds to sampling from the model. Generative models are also known as simulator- or simulation-based models (Hartig et al 2011), or implicit models (Diggle and Gratton 1984), and are closely related to probabilistic programs (Mansinghka et al 2013). Their scope of applicability is extremely wide ranging from genetics and ecology

Michael U. Gutmann
School of Informatics, University of Edinburgh
E-mail: michael.gutmann@ed.ac.uk

Ritabrata Dutta
InterDisciplinary Institute of Data Science, Universitá della Svizzera italiana
E-mail: ritabrata.dutta@usi.ch

Samuel Kaski
Helsinki Institute for Information Technology,
Department of Computer Science, Aalto University
E-mail: samuel.kaski@aalto.fi

Jukka Corander
Department of Biostatistics, University of Oslo
Helsinki Institute for Information Technology,
Department of Mathematics and Statistics, University of Helsinki
E-mail: jukka.corander@medisin.uio.no

---

[1] Early versions were communicated as (Gutmann et al 2014a,b).

(Beaumont 2010) to economics (Gouriéroux et al 1993), physics (Cameron and Pettitt 2012) and computer vision (Zhu et al 2009).

A disadvantage of complex generative models is the difficulty of performing inference with them: evaluating the likelihood function involves computing the probability of the observed data $\mathbf{X}$ as function of the model parameters $\boldsymbol{\theta}$, which for complex models cannot be done analytically or computationally within practical time limits.

As generative models are widely used, solutions have emerged in multiple fields to perform "likelihood-free" inference, that is, inference which does not rely on the availability of the likelihood function. Approximate Bayesian computation (ABC) stems from research in genetics (Beaumont et al 2002; Marjoram et al 2003; Pritchard et al 1999; Tavaré et al 1997), while the method of simulated moments (McFadden 1989; Pakes and Pollard 1989) and indirect inference (Gouriéroux et al 1993; Smith 2008) come from econometrics. The latter methods are traditionally used in a classical inference framework while ABC has its roots in Bayesian inference, but the boundaries have started to blur (Drovandi et al 2011). Despite their differences, the methods all share the basic idea to perform inference about $\boldsymbol{\theta}$ by identifying values which generate simulated data $\mathbf{Y}_{\boldsymbol{\theta}}$ that resemble the observed data $\mathbf{X}$.

The discrepancy between the simulated and observed data is typically measured by reducing each data set to a vector of summary statistics and measuring the distance between them. Both the distance function used and the summary statistics are critical for the success of the inference procedure (see, for example, the reviews by Lintusaari et al 2016; Marin et al 2012). Traditionally, researchers choose the two quantities subjectively, relying on expert knowledge about the observed data. The goal of this paper is to show that the complete arsenal of classification methods can be brought to our disposal to measure the discrepancy, and thus to perform inference for intractable generative models.

The paper is based on the observation that distinguishing two data sets that were generated with very different values of $\boldsymbol{\theta}$ is usually easier than distinguishing two data sets that were generated with similar values. We propose to use the discriminability (classifiability) of the observed and simulated data as a discrepancy measure in likelihood-free inference.

We visualize the basic idea in Figure 1 for the inference of the mean $\boldsymbol{\theta}$ of a bivariate Gaussian with identity covariance matrix. The observed data $\mathbf{X}$, shown with black circles, were generated with mean $\boldsymbol{\theta}^o$ equal to zero. Figure 1a shows that data $\mathbf{Y}_{\boldsymbol{\theta}}$ simulated with mean $\boldsymbol{\theta} = (6, 0)$ can be easily distinguished from $\mathbf{X}$. The indicated classification rule yields an accuracy of 100%. In Figure 1b, on the other hand, the data were simulated with $\boldsymbol{\theta} = (1/2, 0)$ and distinguishing such data from $\mathbf{X}$ is much more difficult; the best classification rule only yields 58% correct assignments. Moreover, if the data were simulated with $\boldsymbol{\theta} = \boldsymbol{\theta}^o$, the classification task could not be solved significantly above chance-level. This suggests that we can perform likelihood-free inference by identifying parameters which yield chance-level discriminability only.

The remaining parts of the paper are structured as follows: In Section 2, we flesh out the basic idea. We then show in Sections 3 and 4 how classification allows us to perform statistical inference of generative models in both a classical and Bayesian framework. The approach will be validated on continuous, binary, discrete, and time series data where ground truth is known. In Section 5, we apply the methodology to real data, and in Section 6, we discuss the proposed approach and related work. Section 7 concludes the paper.

## 2 Measuring discrepancy via classification

Standard classification methods operate on feature vectors that numerically represent the properties of the data that are judged relevant for the discrimination task (Hastie et al 2009; Wasserman 2004). There is some freedom in how the feature vectors are constructed. In the simplest case, the data are statistically independent and identically distributed (iid) random variables, and the features are equal to the data points, as in Figure 1. But the approach of using classification to measure the discrepancy is not restricted to iid data. In the paper, we will construct features and set up a classification problems also for time series or matrix-valued data.

We denote the feature vectors from the observed data $\mathbf{X}$ by $\mathbf{x}_i$, and the feature vectors from the simulated data $\mathbf{Y}_{\boldsymbol{\theta}}$ by $\mathbf{y}_i$, where the dependency on $\boldsymbol{\theta}$ is suppressed for notational simplicity. We assume that we obtained $n$ feature vectors from each of the two data sets. The $\mathbf{x}_i$ are then associated with class label 0 and the $\mathbf{y}_i$ with class label 1, which yields the augmented data set $\mathcal{D}_{\boldsymbol{\theta}}$,

$$\mathcal{D}_{\boldsymbol{\theta}} = \{(\mathbf{x}_1, 0), \ldots, (\mathbf{x}_n, 0), (\mathbf{y}_1, 1), \ldots, (\mathbf{y}_n, 1)\}. \quad (1)$$

Classification consists in predicting the class labels of the features in $\mathcal{D}_{\boldsymbol{\theta}}$. This is done by means of a classification rule $h$ that maps each feature vector $\mathbf{u}$ to its class label $h(\mathbf{u}) \in \{0, 1\}$. The performance of $h$ on $\mathcal{D}_{\boldsymbol{\theta}}$ can be assessed by the classification accuracy CA,

$$\mathrm{CA}(h, \mathcal{D}_{\boldsymbol{\theta}}) = \frac{1}{2n} \left( \sum_{i=1}^{n} [1 - h(\mathbf{x}_i)] + h(\mathbf{y}_i) \right), \quad (2)$$

 

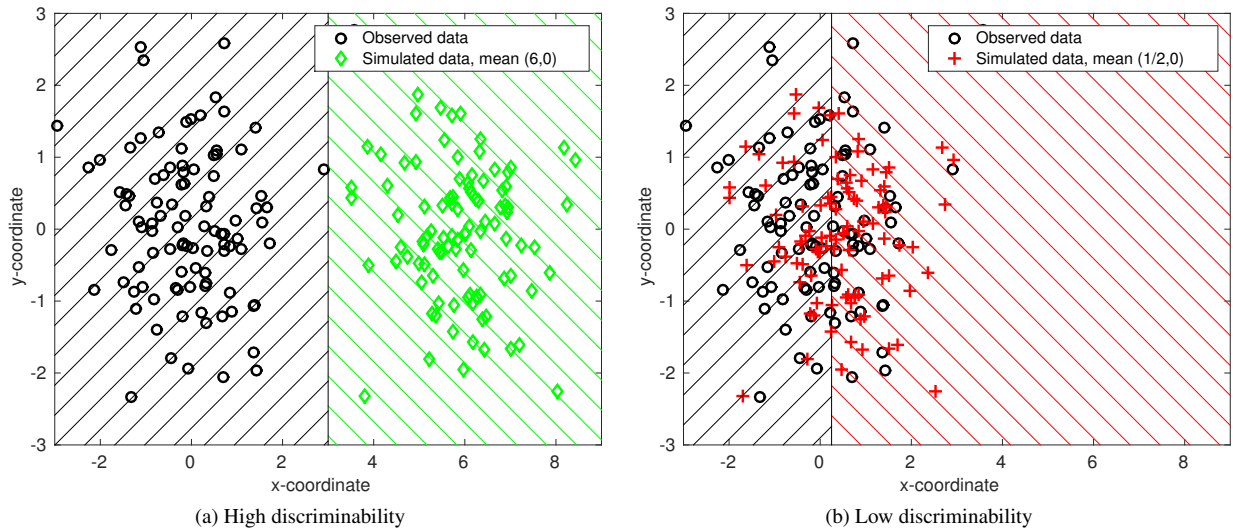

Fig. 1: Discriminability as discrepancy measure. The observed data $\mathbf{X}$ are shown as black circles and were generated with mean $\boldsymbol{\theta}^o = (0, 0)$. The hatched areas indicate the Bayes classification rules. (a) Simulated data $\mathbf{Y}_{\boldsymbol{\theta}}$ (green diamonds) were generated with $\boldsymbol{\theta} = (6, 0)$. (b) $\mathbf{Y}_{\boldsymbol{\theta}}$ (red crosses) were generated with $\boldsymbol{\theta} = (1/2, 0)$. As $\boldsymbol{\theta}$ approaches $\boldsymbol{\theta}^o$, the discriminability (best classification accuracy) of $\mathbf{X}$ and $\mathbf{Y}_{\boldsymbol{\theta}}$ drops. We propose to use the discriminability as discrepancy measure for likelihood-free inference.

which is the proportion of correct assignments. The largest classification accuracy on average is achieved by the Bayes classification rule $h_{\boldsymbol{\theta}}^*$, which consists in assigning a feature vector to $\mathbf{X}$ if it is more probable that the feature belongs to $\mathbf{X}$ than to $\mathbf{Y}_{\boldsymbol{\theta}}$, and vice versa for $\mathbf{Y}_{\boldsymbol{\theta}}$ (Hastie et al 2009; Wasserman 2004). We denote this largest classification accuracy by $J_n^*(\boldsymbol{\theta})$,

$$J_n^*(\boldsymbol{\theta}) = \mathrm{CA}(h_{\boldsymbol{\theta}}^*, \mathcal{D}_{\boldsymbol{\theta}}). \qquad (3)$$

It is an indicator of the discriminability (classifiability) of $\mathbf{X}$ and $\mathbf{Y}_{\boldsymbol{\theta}}$.

In the motivating example in Figure 1, the labels of the data points are indicated by their markers, and the Bayes classification rule by the hatched areas. The classification accuracy $J_n^*(\boldsymbol{\theta})$ decreases from 100% (perfect classification performance) toward 50% (chance-level performance) as $\boldsymbol{\theta}$ approaches $\boldsymbol{\theta}^o$, the parameter value which was used to generate the observed data $\mathbf{X}$. While this provides an intuitive justification for using $J_n^*(\boldsymbol{\theta})$ as discrepancy measure, an analytical justification will be given in the next section where we show that $J_n^*(\boldsymbol{\theta})$ is related to the total variation distance under mild conditions.

In practice, $J_n^*(\boldsymbol{\theta})$ is not computable because the Bayes classification rule $h_{\boldsymbol{\theta}}^*$ involves the probability distribution of the data which is unknown in the first place. But the classification literature provides a wealth of methods to learn an approximation $\hat{h}_{\boldsymbol{\theta}}$ of the Bayes classification rule, and $J_n^*(\boldsymbol{\theta})$ can be estimated via cross-validation (Hastie et al 2009; Wasserman 2004).

We will use several straightforward methods to obtain $\hat{h}_{\boldsymbol{\theta}}$: linear discriminant analysis (LDA), quadratic discriminant analysis (QDA), $L_1$-regularized polynomial logistic regression, $L_1$-regularized polynomial support vector machine (SVM) classification, and an aggregation of the above and other methods (max-rule, see Supplementary material 1.1). These are by no means the only applicable methods. In fact, any method yielding a good approximation of $h_{\boldsymbol{\theta}}^*$ may be chosen; our approach makes the complete arsenal of classification methods available for inference of generative models.

While other approaches are possible, for the approximation of $J_n^*(\boldsymbol{\theta})$, we use $K$-fold cross-validation where the data $\mathcal{D}_{\boldsymbol{\theta}}$ are divided into $K$ folds of training and validation sets, the different validation sets being disjoint. The training sets are used to learn the classification rules $\hat{h}_{\boldsymbol{\theta}}^k$ by any of the methods above, and the validation sets $\mathcal{D}_{\boldsymbol{\theta}}^k$ are used to measure their performances $\mathrm{CA}(\hat{h}_{\boldsymbol{\theta}}^k, \mathcal{D}_{\boldsymbol{\theta}}^k)$. The average classification accuracy on the validation sets, $J_n(\boldsymbol{\theta})$,

$$J_n(\boldsymbol{\theta}) = \frac{1}{K} \sum_{k=1}^{K} \mathrm{CA}(\hat{h}_{\boldsymbol{\theta}}^k, \mathcal{D}_{\boldsymbol{\theta}}^k), \qquad (4)$$

approximates $J_n^*(\boldsymbol{\theta})$, and is used as computable measure of the discrepancy between $\mathbf{X}$ and $\mathbf{Y}_{\boldsymbol{\theta}}$.

We used $K = 5$ folds in the paper. In cross-validation, large values of $K$ generally lead to approximations with smaller bias but larger variance than small values of $K$. Intermediate values like $K = 5$ are thought to lead to a good balance between the two desiderata (e.g. Hastie et al 2009, Section 7.10).

We next show on a range of different kinds of data that most of the different classification methods yield equally good approximations of $J_n^*(\boldsymbol{\theta})$ for large sample sizes. Continuous data (drawn from a univariate Gaussian distribution of variance one), binary data (from a Bernoulli distribution), count data (from a Poisson distribution), and time-series data (from a zero mean moving average model of order one) are considered. For the first three data sets, the unknown parameter is the mean, and for the moving average model, the lag coefficient is the unknown quantity (see Supplementary material 1.2 for the model specifications). Unlike for the other three data sets, the data points from the moving average model are not statistically independent, as the lag coefficient affects the correlation between two consecutive time points $x_t$ and $x_{t+1}$. For the classification, we treated each pair $(x_t, x_{t+1})$ as a feature.

Figure 2 shows that for the Gaussian, Bernoulli, and Poisson data, all the considered classification methods perform as well as the Bayes classification rule (BCR), yielding discrepancy measures $J_n(\boldsymbol{\theta})$ that are practically identical to $J_n^*(\boldsymbol{\theta})$. The same holds for the moving average model, with the exception of LDA. The reason is that LDA is not sensitive to the correlation between $x_t$ and $x_{t+1}$, which would be needed to discover the value of the lag coefficient. In other words, the Bayes classification rule $h_{\boldsymbol{\theta}}^*$ is outside the family of possible classification rules learned by LDA.

The examples show that classification can be used to identify the data generating parameter value $\boldsymbol{\theta}^o$ by minimizing $J_n(\boldsymbol{\theta})$. Further evidence is provided as Supplementary material 2. The derivation of the conditions which guarantee the identification of $\boldsymbol{\theta}^o$ via classification in general is the topic of the next section.

## 3 Classical inference via classification

In this section, we consider the task of finding the single best parameter value. This can be the primary goal of the inference or only the first step before computing the posterior distribution, which will be considered in the following section. In our context, the best parameter value is the value for which the simulated data $\mathbf{Y}_{\boldsymbol{\theta}}$ are the least distinguishable from the observed data $\mathbf{X}$, that is, the parameter $\hat{\boldsymbol{\theta}}_n$ which minimizes $J_n$,

$$\hat{\boldsymbol{\theta}}_n = \operatorname{argmin}_{\boldsymbol{\theta}} J_n(\boldsymbol{\theta}). \qquad (5)$$

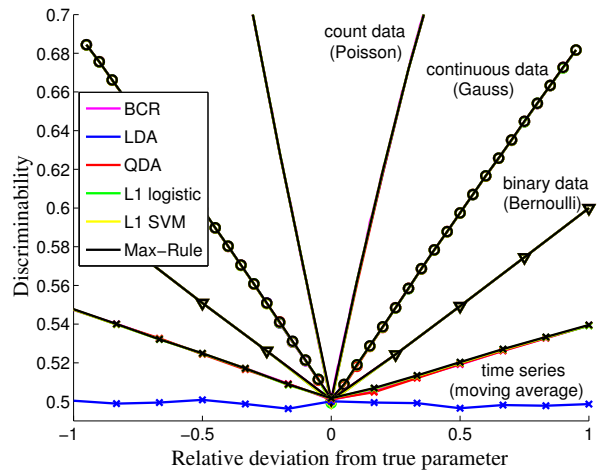

**Fig. 2:** Comparison of the classification accuracy of the Bayes and the learned classification rules for large sample sizes ($n = 100{,}000$). The symmetric curves depict $J_n$ and $J_n^*$ as a function of the relative deviation of the model parameter from the true data generating parameter. As the curves of the different methods are indistinguishable, quadratic discriminant analysis (QDA), $L_1$-regularized polynomial logistic regression (L1 logistic), $L_1$-regularized polynomial support vector machine classification (L1 SVM), and a max-combination of these and other methods (Max-Rule) perform as well as the Bayes classification rule, which assumes the true distributions to be known (BCR). For linear discriminant analysis (LDA), this holds with the exception of the moving average model.

We show that $\hat{\boldsymbol{\theta}}_n$ is a consistent estimator: Assuming that the observed data $\mathbf{X}$ equal some $\mathbf{Y}_{\boldsymbol{\theta}^o}$, generated with unknown parameter $\boldsymbol{\theta}^o$, conditions are given under which $\hat{\boldsymbol{\theta}}_n$ converges to $\boldsymbol{\theta}^o$ in probability as the sample size $n$ increases. Figure 3 provides motivating evidence for consistency of $\hat{\boldsymbol{\theta}}_n$.

The proposition below lists two conditions. The first one is related to convergence of frequencies to expectations (law of large numbers), the second to the ability to learn the Bayes classification rule more accurately as the sample size increases. We prove the proposition in Appendix A. Some basic assumptions are made: The $\mathbf{x}_i$ are assumed to have the marginal probability measure $\mathbf{P}_{\boldsymbol{\theta}^o}$ and the $\mathbf{y}_i$ the marginal probability measure $\mathbf{P}_{\boldsymbol{\theta}}$ for all $i$, which amounts to a weak stationarity assumption. Importantly, the stationarity assumption does not rule out statistical dependencies between the data points; time-series data, for example, are allowed. We also assume that the parametrization of $\mathbf{P}_{\boldsymbol{\theta}}$ is not degenerate, that is, there is a compact set $\Theta$ containing $\boldsymbol{\theta}^o$ where $\boldsymbol{\theta} \neq \boldsymbol{\theta}^o$ implies that $\mathbf{P}_{\boldsymbol{\theta}} \neq P_{\boldsymbol{\theta}^o}$.

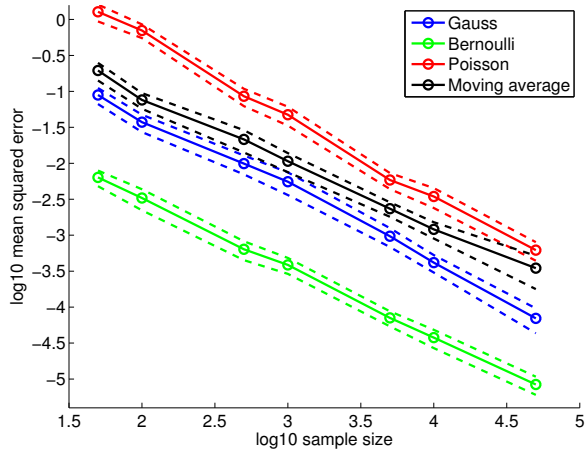

Fig. 3: Empirical evidence for consistency. The figure shows the mean squared estimation error $\mathrm{E}[||\hat{\boldsymbol{\theta}}_n - \boldsymbol{\theta}^o||^2]$ for the examples in Figure 2 as a function of the sample size $n$ (solid lines, circles). The mean was computed as an average over 100 outcomes. The dashed lines depict the mean $\pm$ 2 standard errors. The linear trend on the log-log scale suggests convergence in quadratic mean, and hence consistency of the estimator $\hat{\boldsymbol{\theta}}_n$. The results are for $L_1$-regularized logistic regression, see Supplementary material 3 for the other classification methods.

**Proposition 1** *Denote the set of features which the Bayes classification rule $h_{\boldsymbol{\theta}}^*$ classifies as being from the simulated data by $H_{\boldsymbol{\theta}}^*$. The expected discriminability $\mathrm{E}(J_n^*(\boldsymbol{\theta}))$ equals $J(\boldsymbol{\theta})$,*

$$J(\boldsymbol{\theta}) = \frac{1}{2} + \frac{1}{2}\left(\mathrm{P}_{\boldsymbol{\theta}}(H_{\boldsymbol{\theta}}^*) - \mathrm{P}_{\boldsymbol{\theta}^o}(H_{\boldsymbol{\theta}}^*)\right), \qquad (6)$$

*and $\hat{\boldsymbol{\theta}}_n$ converges to $\boldsymbol{\theta}^o$ in probability as the sample size $n$ increases, $\hat{\boldsymbol{\theta}}_n \xrightarrow{P} \boldsymbol{\theta}^o$, if*

$$\sup_{\boldsymbol{\theta} \in \Theta} |J_n^*(\boldsymbol{\theta}) - J(\boldsymbol{\theta})| \xrightarrow{P} 0 \quad and \qquad (7)$$

$$\sup_{\boldsymbol{\theta} \in \Theta} |J_n(\boldsymbol{\theta}) - J_n^*(\boldsymbol{\theta})| \xrightarrow{P} 0. \qquad (8)$$

The two conditions guarantee that $J_n(\boldsymbol{\theta})$ converges uniformly to $J(\boldsymbol{\theta})$, so that $J(\boldsymbol{\theta})$ is minimized with the minimization of $J_n(\boldsymbol{\theta})$ as $n$ increases. Since $J(\boldsymbol{\theta})$ attains its minimum at $\boldsymbol{\theta}^o$, $\hat{\boldsymbol{\theta}}_n$ converges to $\boldsymbol{\theta}^o$. By definition of $H_{\boldsymbol{\theta}}^*$, $\mathrm{P}_{\boldsymbol{\theta}}(H_{\boldsymbol{\theta}}^*) - \mathrm{P}_{\boldsymbol{\theta}^o}(H_{\boldsymbol{\theta}}^*)$ is one-half of the total variation distance between the two distributions (Pollard 2001, Chapter 3). The limiting objective $J(\boldsymbol{\theta})$ corresponds thus to a well defined statistical distance between $\mathrm{P}_{\boldsymbol{\theta}}$ and $\mathrm{P}_{\boldsymbol{\theta}^o}$.

The condition in Equation (7) is about convergence of sample averages to expectations. Standard convergence results apply for statistically independent features. For features with statistical dependencies, like for example for time-series data, corresponding convergence results are investigated in empirical process theory (van der Vaart and Wellner 1996), which forms a natural limit of what is studied in this paper. We may only note that by definition of $J$, convergence will depend on the complexity of the sets $H_{\boldsymbol{\theta}}^*$, $\boldsymbol{\theta} \in \Theta$, and hence the complexity of the Bayes classification rules $h_{\boldsymbol{\theta}}^*$. The condition does not depend on the classification method employed. In other words, this first condition is about the difficulty of the classification problems that need to be solved. The condition in Equation (8), on the other hand, is about the ability to solve them: The performance of the learned rule needs to approach the performance of the Bayes classification rule as the number of available samples increases. How to best learn such rules and finding conditions which guarantee successful learning is a research area in itself (Zhang 2004).

In Figure 2, LDA did not satisfy the condition in Equation (8) for the moving average data, which can be seen by the chance-level performance for all parameters tested. This failure of LDA suggests a practical means to test whether the second condition holds: We generate data sets with two very different parameter values so that it is unlikely that the data sets are similar to each other, and learn to discriminate between them. If the performance is persistently close to chance-level, the Bayes classification rule is likely outside the family of classification rules that the method is able to learn, so that the condition would be violated. Regarding the first condition, the results in Figure 3 suggest that it is satisfied for all four inference problems considered. Generally verifying whether the sample average converges to the expectation, e.g. via a general method that works reliably for any kind of time-series data, seems, however, difficult.

## 4 Bayesian inference via classification

We consider next inference of the posterior distribution of $\boldsymbol{\theta}$ in the framework of approximate Bayesian computation (ABC).

ABC comprises several simulation-based methods to obtain samples from the posterior distribution when the likelihood function is not known (for review papers, see, for example, Lintusaari et al 2016; Marin et al 2012). ABC algorithms are iterative: The basic steps at each iteration are

1. proposing a parameter value $\boldsymbol{\theta}'$,
2. simulating pseudo observed data $\mathbf{Y}_{\boldsymbol{\theta}'}$, and then
3. accepting or rejecting the proposal based on a comparison of $\mathbf{Y}_{\boldsymbol{\theta}'}$ with the real observed data $\mathbf{X}$.

How to actually measure the discrepancy between the observed and the simulated data is a major difficulty in

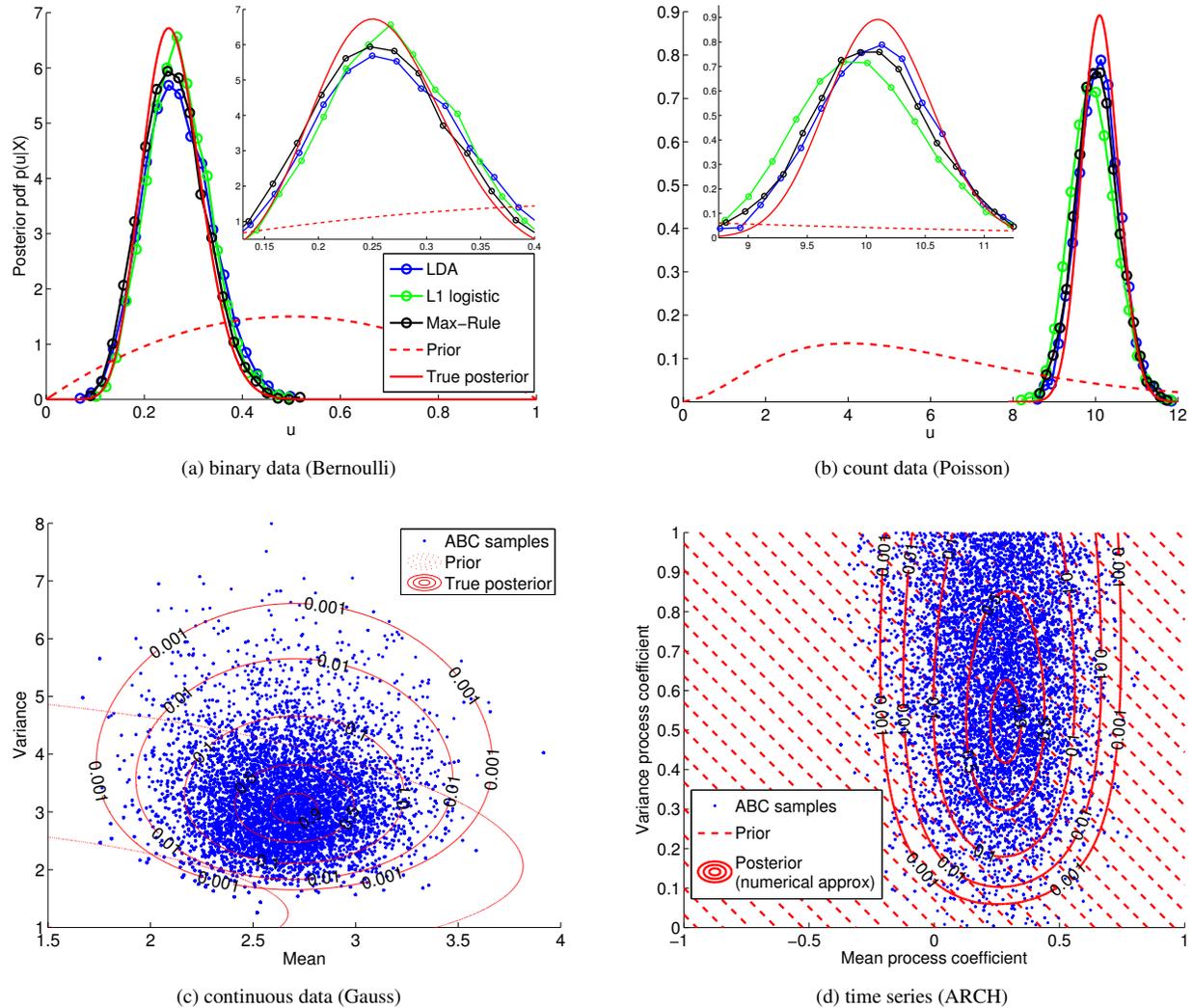

(a) binary data (Bernoulli)

(b) count data (Poisson)

(c) continuous data (Gauss)

(d) time series (ARCH)

Fig. 4: Posterior distributions inferred by classifier ABC for binary, count, continuous and time-series data. The results are for 10,000 ABC samples and $n = 50$. For the univariate cases, the samples are summarized as empirical pdfs. For the bivariate cases, scatter plots of the obtained samples are shown (the results are for the max-rule). The numbers on the contours are relative to the maximum of the reference posterior. For the autoregressive conditional heteroskedasticity (ARCH) model, the hatched area indicates the domain of the uniform prior. Supplementary material 4 contains additional examples and results.

these methods (Lintusaari et al 2016; Marin et al 2012). We here show that $J_n$ can be used as a discrepancy measure in ABC; in the following, we call this approach "classifier ABC." In step 3, we thus compare $\mathbf{Y}_{\boldsymbol{\theta}'}$ and $\mathbf{X}$ through the lenses of a classifier by computing the discriminability of the two data sets.

The results reported in this paper were obtained with a sequential Monte Carlo implementation (see Supplementary material 1.3). The use of $J_n$ in ABC is, however, not restricted to that particular algorithm.

We validated classifier ABC on binary (Bernoulli), count (Poisson), continuous (Gaussian), and time-series (ARCH) data (see Supplementary material 1.2 for the model details). The true posterior for the autoregressive conditional heteroskedasticity (ARCH) model is not available in closed form. We approximated it using deterministic numerical integration, as detailed in Supplementary material 1.2.

The inferred empirical posterior probability density functions (pdfs) are shown in Figure 4. There is a good match with the true posterior pdfs or the approxima-tion obtained with deterministic numerical integration. Different classification methods yield different results but the overall performance is rather similar. Regarding computation time, the simpler LDA and QDA tend to be faster than the other classification methods used, with the max-rule being the slowest one. Additional examples as well as links to movies showing the evolution of the posterior samples in the ABC algorithm can be found in Supplementary material 4.

As a quantitative analysis, we computed the relative error of the posterior means and standard deviations. The results, reported as part of Supplementary material 4, show that the errors in the posterior mean are within 5% after five iterations of the ABC algorithm for the examples with independent data points. For the time series, where the data points are not independent, a larger error of 15% occurs. The histograms and scatter plots show, however, that the corresponding ABC samples are still very reasonable.

## 5 Application on real data

We next used our approach to infer an intractable model of bacterial infections in child care centers.

### 5.1 Data and model

The observed data $\mathbf{X}$ were the presence or absence of different strains of the bacterium *Streptococcus pneumoniae* among attendees of $M = 29$ child care centers in the metropolitan area of Oslo, Norway, at single points of time $T_m$ (cross-sectional data). On average, $N = 53$ children attended a center. Only a subset of size $N_m$ of all attendees of each center was sampled. The data were collected and first described by Vestrheim et al (2008).

In the following, we represent the colonization state of individual $i$ in a child care center by the binary variable $I_{is}^t, s = 1, \ldots, S$, where $S$ the total number of strains in circulation. If the attendee is infected with strain $s$ of the bacterium at time $t$, $I_{is}^t = 1$, and otherwise, $I_{is}^t = 0$. The observed data $\mathbf{X}$ consisted thus of a set of $M = 29$ binary matrices of size $N_m \times S$ formed by the $I_{is}^{T_m}, i = 1, \ldots, N_m, s = 1, \ldots, S$.

The model for which we performed inference was developed by Numminen et al (2013). It is individual-based and consists of a continuous-time Markov chain for the transmission dynamics inside a child care center paired with an observation model. The child care centers were assumed independent. The model is sketched in Figure 5 for a single center.

In each child care center, the transmission dynamics started with zero infected individuals, $I_{is}^0 = 0$ for all $i$ and $s$, after which the states evolved in a stochastic

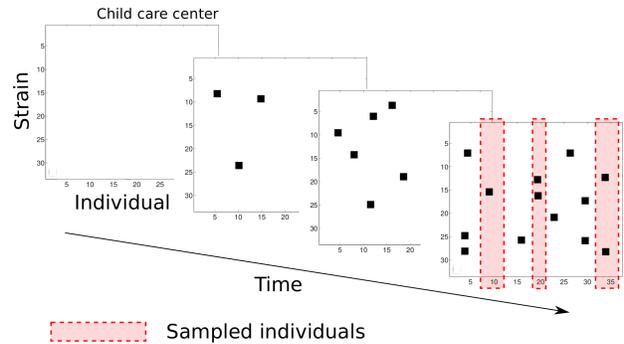

Fig. 5: Sketch of the individual-based epidemic model. The evolution of the colonization states in a single child care center is shown. Colonization is indicated by the black squares.

manner according to the following transition probabilities:

$$P(I_{is}^{t+h} = 0 | I_{is}^t = 1) = h + o(h), \tag{9}$$

$$P(I_{is}^{t+h} = 1 | I_{is'}^t = 0 \; \forall s') = R_s^t h + o(h), \tag{10}$$

$$P(I_{is}^{t+h} = 1 | I_{is}^t = 0, \; \exists s' : I_{is'}^t = 1) = \theta R_s^t h + o(h), \tag{11}$$

where $h$ is a small time interval and $o(h)$ a remainder term satisfying $\lim_{h \to 0} o(h)/h = 0$. Equation (9) describes the probability to clear strain $s$, Equation (10) the probability to be infected by it when previously not infected with any strain, and Equation (11) the probability to be infected by it when previously infected with another strain $s'$. The rate of infection with strain $s$ at time $t$ is denoted by $R_s^t$, and $\theta \in (0, 1)$ is an unknown co-infection parameter. For $\theta = 0$, the probability for a co-infection is zero. The rate $R_s^t$ was modeled as

$$R_s^t = \beta E_s^t + \Lambda P_s, \tag{12}$$

$$E_s^t = \sum_{j=1}^{N} \frac{1}{N-1} \frac{I_{js}^t}{n_j^t}, \tag{13}$$

$$n_j^t = \sum_{s'=1}^{S} I_{js'}^t, \tag{14}$$

where $N$ is the average number of children attending the child care center, and $\Lambda$ and $\beta$ are two unknown rate parameters that scale the static probability $P_s$ for an infection happening outside the child care center and the dynamic probability $E_s^t$ for an infection from within, respectively. The probability $P_s$ and the number of strains $S$ were determined by an analysis of the overall distribution of the strains in the cross-sectional data (yielding $S = 33$; for $P_s$, see Numminen et al 2013). The expression for $E_s^t$ in Equation (13) was derived by assuming that contacts happen uniformly at random (the probability for a contact is $1/(N-1)$), and that the strains attendee $j$ is carrying are all transmitted with equal probability (with $n_j^t$ being the total number of strains carried by attendee $j$, the probability for a transmission of strain $s$ is $I_{js}^t/n_j^t$).

The observation model was random sampling of $N_m$ individuals without replacement from the average number $N$ of individuals attending a child care center. A stationarity assumption was made so that the exact value of the sampling time $T_m$ was not of importance as long as it is sufficiently large so that the system is in its stationary regime.

The model has three parameters for which uniform priors were assumed: Parameter $\beta \in (0, 11)$ which is related to the probability to be infected by someone inside a child care center, parameter $\Lambda \in (0, 2)$ for the probability of an infection from an outside source, and parameter $\theta \in (0, 1)$ which is related to the probability to be infected with multiple strains. With a slight abuse of notation, we will use $\boldsymbol{\theta} = (\beta, \Lambda, \theta)$ to denote the compound parameter vector.

### 5.2 Reference inference method

Since the likelihood function is intractable, the model was inferred with ABC in previous work (Numminen et al 2013). The summary statistics were chosen based on epidemiological considerations and the distance function was adapted to the specific problem at hand.

To compare $\mathbf{X}$ and $\mathbf{Y}_{\boldsymbol{\theta}}$, Numminen et al (2013) first summarized each of the $M = 29$ child care centers of the simulated and observed data using four statistics,

1. the strain diversity in the child care centers,
2. the number of different strains circulating,
3. the proportion of individuals who are infected, and
4. the proportion of individuals who are infected with more than one strain.

For each of the four summary statistics, the empirical cumulative distribution function (cdf) was computed from the obtained $M = 29$ values. The $L_1$ distances between the empirical cdfs of the summary statistics for $\mathbf{X}$ and $\mathbf{Y}_{\boldsymbol{\theta}}$ were then used to assess the discrepancy (Numminen et al 2013). Inference was performed with a sequential Monte Carlo ABC algorithm with four generations. The corresponding posterior distribution will serve as reference against which we compare the solution by classifier ABC.

### 5.3 Formulation as classification problem

For likelihood-free inference via standard classification, the observed matrix-valued data were transformed to feature vectors. We used simple features which reflect the matrix structure and the binary nature of the data.

For the matrix-nature of the data, the rank of each matrix and the $L_2$-norm of the singular values (scaled by the size of the matrix) were used. For the binary nature of the data, we counted the fraction of ones in certain subsets of each matrix and used the average of the counts and their variability as features. The set of rows and the set of columns were used, as well as 100 randomly chosen subsets. Each random subset contained 10% of the elements of a matrix. Since the average of the counts is the same for the row and column subsets (it equals the fraction of all ones in a matrix), only one average was used.

The features $\mathbf{x}_i$ or $\mathbf{y}_i$ in the classification had thus size seven (2 dimensions are for the matrix properties, 3 dimensions for the column and row subsets, and 2 dimensions for the random subsets). Multiple random subsets can be extracted from each matrix. We made use of this to obtain $n = 1{,}000$ features $\mathbf{x}_i$ and $\mathbf{y}_i$. We also ran classifier ABC without random subsets; the classification problems consisted then in discriminating between two data sets consisting each of 29 five-dimensional feature vectors. As classification method, we used LDA.

### 5.4 Inference results

In ABC, the applicability of a discrepancy measure can be assessed by first performing inference on synthetic data of the same size and structure as the observed data but simulated from the model with known parameter values. Since ABC algorithms are rather time consuming, we first tested the applicability of $J_n$ in the framework of point estimation. We computed $J_n(\boldsymbol{\theta})$ varying only two of the three parameters at a time, keeping the third parameter fixed at the value which was used to generate the data. To eliminate random effects, we used for all $\boldsymbol{\theta}$ the same random number generator seed when simulating the $\mathbf{Y}_{\boldsymbol{\theta}}$. The seeds for $\mathbf{X}$ and the $\mathbf{Y}_{\boldsymbol{\theta}}$ were different.

Figure 6 shows the results for classification with randomly chosen subsets (top row) and without (bottom row). The diagrams on the top and bottom row are very similar, both have well-defined regions in the parameter space for which $J_n$ is close to one half, which corresponds to chance-level discriminability. But the features from the random subsets were helpful to discriminate between $\mathbf{X}$ and $\mathbf{Y}_{\boldsymbol{\theta}}$ and produced more localized regions with small $J_n$. The results suggest that LDA, the arguably simplest classification method, is suitable to infer the epidemic model.

We next applied classifier ABC on the synthetic data, using a sequential Monte Carlo ABC algorithm with four generations as previously done by Numminen et al (2013).

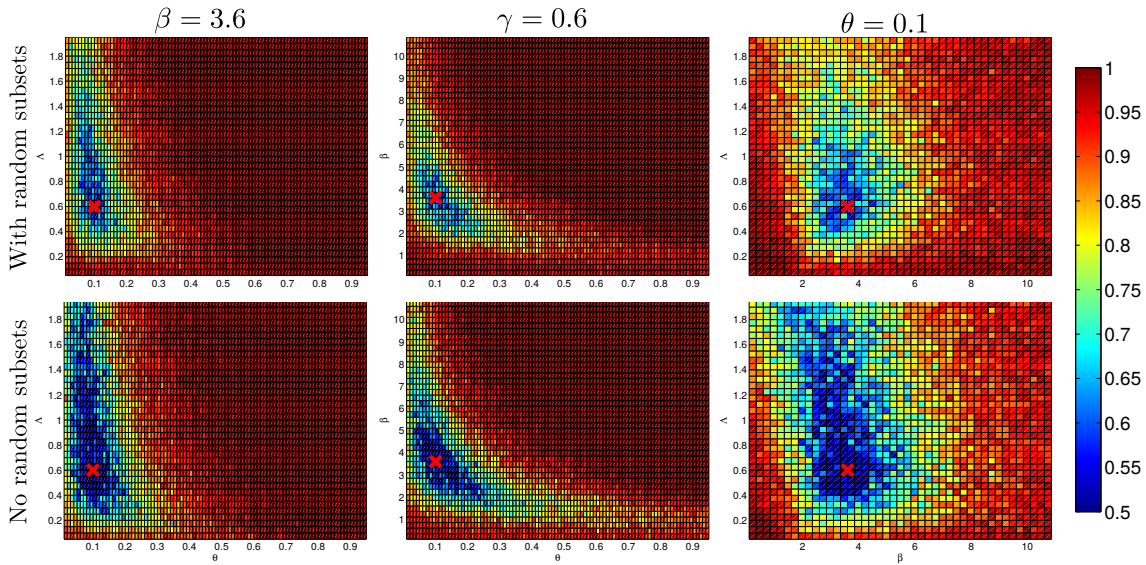

Fig. 6: Testing the applicability of the discrepancy measure $J_n$ to infer the individual-based epidemic model. The figures show $J_n(\boldsymbol{\theta})$ when one parameter is fixed at a time. The red crosses mark the data generating parameter value $\boldsymbol{\theta}^o = (\beta^o, \Lambda^o, \theta^o) = (3.6, 0.6, 0.1)$. The presence of random features produced more localized regions with small $J_n$.

The resulting posterior pdfs are shown in Figure 7 in form of kernel density estimates (smoothed and scaled histograms) based on 1,000 ABC samples. It can be seen that classifier ABC with or without random subsets both yielded results which are qualitatively similar to the expert solution. The strongest difference is that the tails of the posterior pdf of $\beta$ are heavier for classifier ABC than for the expert solution. In case of classifier ABC with random subsets, this difference became less pronounced when the algorithm was run for an additional fifth iteration (Supplementary material 5). For classifier ABC without random subsets, on the other hand, the difference persisted. This behavior is in line with Figure 6 where the random features led to tighter $J_n$-diagrams. Overall, the results on synthetic data confirm the applicability of classifier ABC to infer the epidemic model.

The results on real data are shown in Figure 8. It can be seen that the posterior distributions obtained with classifier ABC are generally similar to the expert solution. The posterior mode of $\beta$ for classifier ABC with random subsets is slightly smaller than for the other methods. The shift could be due to stochastic variation because we only worked with 1,000 ABC samples. It could, however, also be that the random features picked up some properties of the real data which the other methods are not sensitive to.

The computation time of classifier ABC with LDA was about the same as for the method by Numminen et al (2013): On average, the total time for the data gen-eration and the discrepancy measurement was $28.49 \pm 3.45$ seconds for LDA while it was $28.41 \pm 3.45$ seconds for the expert method; with $28.4 \pm 3.45$ seconds, most of the time was spent on generating data from the epidemic model. Altogether, classifier ABC thus yielded inference results which are equivalent to the expert solution, from both a statistical and computational point of view.

### 5.5 Compensating for missing expert statistics

So far we did not use expert-knowledge about the inference problem when solving it with classifier ABC. Using discriminability in a classification task as a discrepancy measure is a data-driven approach to assess the similarity between simulated and observed data. But it is not necessarily a black-box approach. Knowledge about the problem at hand can be incorporated when specifying the classification problem. Furthermore, the approach is compatible with summary statistics derived from expert knowledge: Classifier ABC, and more generally the discrepancy measure $J_n$, is able to incorporate the expert statistics by letting them be features (covariates) in the classification. The combined use of expert statistics and classifier ABC enables one to filter out properties of the model which are either not of interest or known to be wrong. Moreover, it makes the inference more robust, for example to possible misspecifications or insufficiencies of the summary statistics, as we illustrate next.

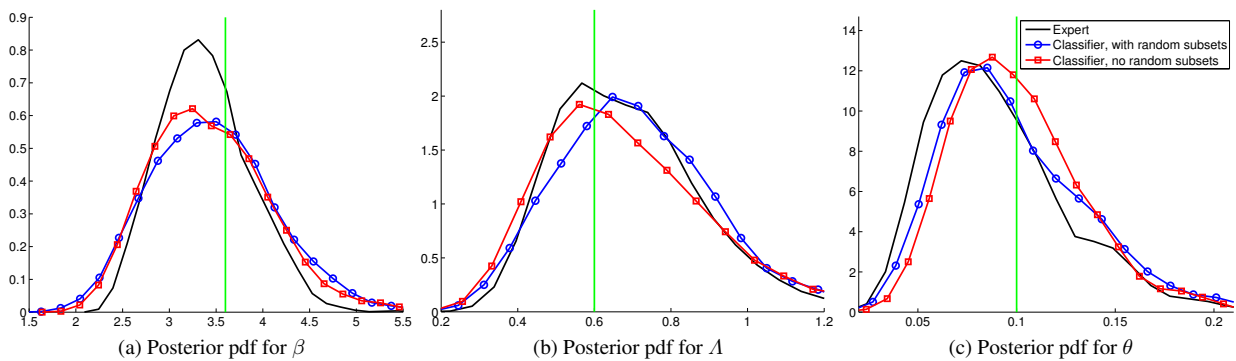

Fig. 7: Inferring the individual-based epidemic model with classifier ABC. The results are for simulated data with known data generating parameter $\theta^o$ (indicated by the green vertical lines). Classifier ABC with random subsets (blue, circles) or without (red, squares) both yielded posterior pdfs which are qualitatively similar to the expert solution (black).

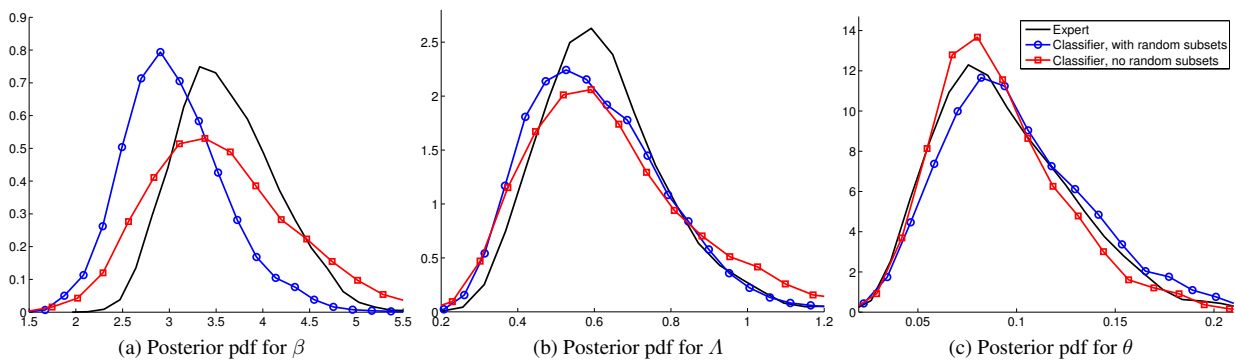

Fig. 8: Inference results on real data, visualized as in Figure 7.

We selected two simple expert statistics used by Numminen et al (2013), namely the number of different strains circulating and the proportion of infected individuals, and inferred the posteriors with this reduced set of summary statistics, using the method by Numminen et al (2013) as before. Figure 9 shows that consequently, the posterior distributions of $\Lambda$ and $\theta$ deteriorated. The used expert statistics alone were insufficient to perform ABC. Combining the insufficient set of summary statistics with classifier ABC, however, led to a recovery of the posteriors. The result are for classifier ABC with random subsets, but the same holds for classifier ABC without random subsets (Supplementary material 5).

## 6 Discussion

Generative models are useful and widely applicable for dealing with uncertainty and for making inferences from data. The intractability of the likelihood function is, however, often a serious problem in the inference for realistic models. While likelihood-free methods provide a powerful framework for performing inference, a limiting difficulty is the required discrepancy measurement between simulated and observed data.

We found that classification can be used to measure the discrepancy. This finding has practical value because it reduces the difficult problem of choosing an appropriate discrepancy measure to a more standard problem where we can leverage a wealth of existing solutions; whenever we can classify, we can do likelihood-free inference. It offers also theoretical value because it reveals that classification can yield consistent likelihood-free inference, and that the two fields of research, which appear very much separated at first glance, are actually tightly connected.

### 6.1 Summary statistics versus features

In the proposed approach, instead of choosing summary statistics and a distance function between them as in the standard approach, we need to choose a classification method and the features. The reader may thus wonder whether we replaced one possibly arbitrary choice with another. The important point is that by choosing a classification method, we only decide about a function space, and not the classification rule itself. The classification rule that is finally used to measure the discrepancy is learned from data and is not specified by the user, which is in stark contrast to the traditional approach based on fixed summary statistics. Moreover, the function space can be chosen using cross-validation, as implemented with our max-rule, which reduces the arbitrariness even more. In Figure Figure 2, for example, the max-rule successfully chose to use other classification methods than LDA for the inference of the moving average model. The influence of the choice of features is also rather mild, because they only affect the discrepancy measurement via the learned classification rule. This property of the proposed approach allowed us to even use random features in the inference of the epidemic model.

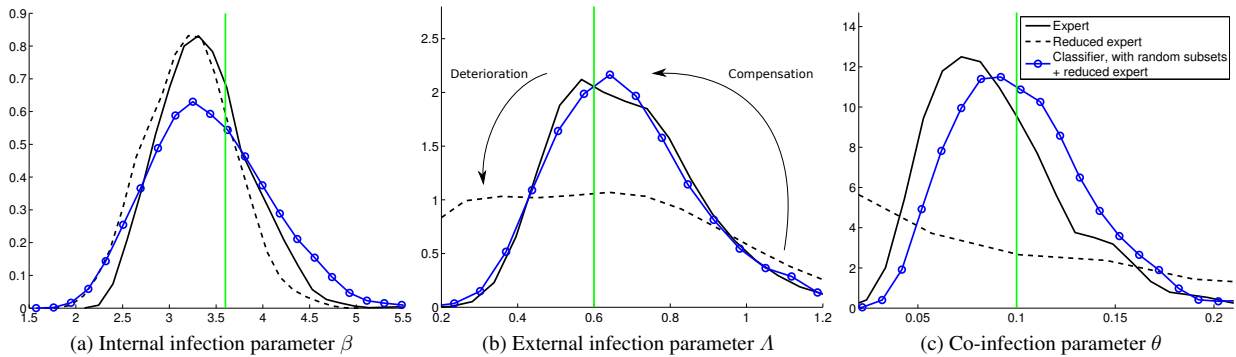

**Fig. 9:** Using classifier ABC to compensate for insufficient expert statistics. The setup and visualization is as in Figure 7. Its expert solution is reproduced for reference. Working with a reduced set of expert statistics affects the posteriors of $\Lambda$ and $\theta$ adversely but classifier ABC is able to compensate (blue curves with circles versus black dashed curves).

The possibility to use random features, however, does not mean that we should not use reliable expert knowledge when available. Indeed, summary statistics derived from expert knowledge can be included by letting them be features (covariates) in the classification.

### 6.2   Related work

In previous work, regression with the parameters $\boldsymbol{\theta}$ as response variables was used to generate summary statistics from a larger pool of candidates (Aeschbacher et al 2012; Fearnhead and Prangle 2012; Wegmann et al 2009). The shared characteristic of these works and our approach is the learning of transformations of the summary statistics and the features, respectively. The criteria which drive the learning are, however, rather different.

Since the candidate statistics are a function of the simulated data $\mathbf{Y}_{\boldsymbol{\theta}}$, we may consider the regression to provide an approximate inversion of the data generation process $\boldsymbol{\theta} \mapsto \mathbf{Y}_{\boldsymbol{\theta}}$. In this interpretation, the (Euclidean) distance of the summary statistics is an approximation of the (Euclidean) distance of the parameters. The optimal inversion of the data generating process in a mean squared error sense is the conditional expectation $\mathrm{E}(\boldsymbol{\theta}|\mathbf{Y}_{\boldsymbol{\theta}})$. Fearnhead and Prangle (2012) showed that this conditional expectation is also the optimal summary statistic for $\mathbf{Y}_{\boldsymbol{\theta}}$ if the goal is to infer $\boldsymbol{\theta}^o$ as accurately as possible under a quadratic loss. Transformations based on regression are thus strongly linked to the computation of the distance between the parameters. The reason we learn transformations, on the other hand, is that we would like to approximate $J_n^*(\boldsymbol{\theta})$ well, which is linked to the computation of the total variation distance between the distributions indexed by the parameters.

Classification was recently used in other work on ABC, but in a different manner. Intractable density ratios in Markov chain Monte Carlo algorithms were estimated using tools from classification (Pham et al 2014), in particular random forests, and Pudlo et al (2016) used random forests for model selection by learning to predict the model class from the simulated data instead of computing their posterior probabilities. This is different from using classification to define a discrepancy measure between simulated and observed data, as done here.

A particular classification method, (nonlinear) logistic regression, was used for the estimation of unnormalized models (Gutmann and Hyvärinen 2012), which are models where the probability density functions are known up to the normalizing partition function only (see Gutmann and Hyvärinen (2013a) for a review paper, and Barthelmé and Chopin (2015); Gutmann and Hirayama (2011); Pihlaja et al (2010) for generalizations). Likelihood-based inference is intractable for unnormalized models but unlike in the generative models considered here, the shape of the model-pdf is known which can be exploited in the inference.

At about the same time we first presented our work (Gutmann et al 2014a,b), Goodfellow et al (2014) proposed to use nonlinear logistic regression to train a neural network such that it transforms "noise" samples into samples approximately following the same distribution as some given data set. The main difference to our work is that the method of Goodfellow et al (2014) is a method for producing random samples while ours is a method for statistical inference.

### 6.3 Sequential inference and prediction

We did not make any specific assumptions about the model or the structure of the observed data $\mathbf{X}$. An interesting special case occurs when $\mathbf{X}$ are an element $\mathbf{X}^{(t_0)}$ of a sequence of data sets $\mathbf{X}^{(t)}$ which are observed one after the other, and the generative model is specified accordingly to generate a sequence of simulated data sets.

For inference at $t_0$, we can distinguish between simulated data which were generated either before or after $\mathbf{X}^{(t_0)}$ are observed: In the former case, the simulated data are predictions about $\mathbf{X}^{(t_0)}$, and after observation of $\mathbf{X}^{(t_0)}$, likelihood-free inference about $\boldsymbol{\theta}$ corresponds to assessing the accuracy of the predictions. That is, the discrepancy measurement converts the predictions of $\mathbf{X}^{(t_0)}$ into inferences of the causes of $\mathbf{X}^{(t_0)}$. In the latter case, each simulated data set can immediately be compared to $\mathbf{X}^{(t_0)}$ which enables efficient iterative identification of parameter values with low discrepancy (Gutmann and Corander 2016). That is, the possible causes of $\mathbf{X}^{(t_0)}$ can be explained more accurately with the benefit of hindsight.

### 6.4 Relation to perception and artificial intelligence

Probabilistic modeling and inference play key roles in image understanding (Gutmann and Hyvärinen 2013b), robotics (Thrun et al 2006), and artificial intelligence (Ghahramani 2015). Perception has been modeled as (Bayesian) inference based on a "mental" generative model of the world (e.g. Vincent 2015). In most of the literature, variational approximate inference has been used for intractable generative models, giving rise to the Helmholtz machine (Dayan et al 1995) and to the free-energy in neuroscience (Friston 2010). But other approximate inference methods can be considered as well.

The discussion about sequential inference and prediction points to similarities between perception and likelihood-free inference or approximate Bayesian computation. It is intuitively sensible that perception would involve prediction of new sensory input given the past, as well as an assessment of the predictions and a refinement of their explanations after arrival of the data. The quality of the inference depends on the quality of the generative model and the quality of the discrepancy assessment. That is, the inference results may only be useful if the generative model of the world is rich enough to produce data resembling the observed data, and if the discrepancy measure can reliably distinguish between the "mentally" generated and the actually observed data.

We proposed to measure the discrepancy via classification, being agnostic about the particular classifier used. It is an open question how to generally best measure the classification accuracy when the data are arriving sequentially. Classifiers are, however, rather naturally part of perceptual systems. Rapid object recognition, for instance, can be achieved via feedforward multilayer classifiers (Serre et al 2007) and there are several techniques to learn representations which facilitate classification (Bengio et al 2013). It is thus conceivable that a given classification machinery is used for several purposes, for example to quickly recognize certain objects but also to assess the discrepancy between simulated and observed data.

## 7 Conclusions and future work

In the paper, we proposed to measure the discrepancy in likelihood-free inference via classification. We focused on the principle and not on a particular classification method. Some methods may be particularly suited for certain models, where it may be possible to measure the discrepancy via the loss function that is used to learn the classification rule instead of the classification accuracy.

When working with the classification accuracy, we only use a single bit of information per data point. While this is little information, we showed that the approach yielded accurate posterior inferences and that it defines a consistent estimator. The Bayesian inference results were empirical and it is likely that a more rigorous theoretical analysis will reveal that the single bit of information puts a limit on the possible closeness to the true posterior. While our empirical results suggest that other error sources may be more dominant in practice, the bottleneck can be avoided by using the current setup to identify the relevant summary statistics, or some transformation of them, and by computing the discrepancy by their Euclidean distance as in classical ABC. While this is a possible approach, in recent work, we chose another path by training the classifier on two simulated data sets whose size can be made as large as computationally possible (Dutta et al 2016).

We here worked with a single simulated data set per parameter value. If multiple simulated data sets are available, they may be used to define an approximate likelihood function by, for example, averaging their corresponding discrepancies (see e.g. Gutmann and Corander 2016, Section 3.3). The approximate likelihood function can then be maximised with respect to the parameters or used in place of the actual likelihood function in standard methods for posterior sampling.

Further exploration of the connection between classification and likelihood-free inference is likely to lead to practical improvements in general: Each parameter $\boldsymbol{\theta}$, for instance, induces a classification problem. We here treated the classification problems separately but they are actually related. First, the observed data $\mathbf{X}$ occur in all the classification problems. Second, the simulated data sets $\mathbf{Y}_{\boldsymbol{\theta}}$ are likely to share some properties if the parameters are not too different. Taking advantage of the relation between the different classification problems may lead to both computational and statistical gains. In the classification literature, leveraging the solution of one problem to solve another one is generally known as transfer learning (Pan and Yang 2010). In the same spirit, leveraging transfer learning, or other methods from classification, seems promising to further advance likelihood-free inference.

**Acknowledgements** The work was partially done when MUG and RD were at the Department of Mathematics and Statistics, University of Helsinki, and the Department of Computer Science, Aalto University, respectively. The work was supported by ERC grant no. 239784 and the Academy of Finland (Finnish Centre of Excellence in Computational Inference Research COIN). RD is presently funded by Swiss National Science Foundation grant *no.*105218_163196. We thank Elina Numminen for providing computer code for the epidemic model.

# Appendix A    Proof of proposition 1

Proposition 1 is proved using an approach based on uniform convergence in probability of $J_n$ to a function $J$ whose minimizer is $\boldsymbol{\theta}^o$ (van der Vaart 1998). The proof has three steps: First, we identify $J$. Second, we find conditions under which $J$ is minimized by $\boldsymbol{\theta}^o$. Third, we derive conditions which imply that $J_n$ converges to $J$.

## A.1    Definition of $J$

For validation sets $\mathcal{D}_{\boldsymbol{\theta}}^k$ consisting of $2m$ labeled features $(\mathbf{x}_i^k, 0)$ and $(\mathbf{y}_i^k, 1)$, $i = 1, \ldots, m$, we have by definition of $\mathrm{CA}(h, \mathcal{D}_{\boldsymbol{\theta}})$ in Equation (2)

$$\mathrm{CA}(\hat{h}_{\boldsymbol{\theta}}^k, \mathcal{D}_{\boldsymbol{\theta}}^k) = \frac{1}{2m}\left(\sum_{i=1}^m [1 - \hat{h}_{\boldsymbol{\theta}}^k(\mathbf{x}_i^k)] + \hat{h}_{\boldsymbol{\theta}}^k(\mathbf{y}_i^k)\right) \quad (15)$$

$$= \frac{1}{2} + \frac{1}{2m}\sum_{i=1}^m \hat{h}_{\boldsymbol{\theta}}^k(\mathbf{y}_i^k) - \hat{h}_{\boldsymbol{\theta}}^k(\mathbf{x}_i^k), \quad (16)$$

so that $J_n(\boldsymbol{\theta})$ in Equation (4) can be written as

$$J_n(\boldsymbol{\theta}) = \frac{1}{K}\sum_{k=1}^K \left(\frac{1}{2} + \frac{1}{2m}\sum_{i=1}^m \hat{h}_{\boldsymbol{\theta}}^k(\mathbf{y}_i^k) - \hat{h}_{\boldsymbol{\theta}}^k(\mathbf{x}_i^k)\right) \quad (17)$$

$$= \frac{1}{2} + \frac{1}{2Km}\sum_{i=1}^m \sum_{k=1}^K \hat{h}_{\boldsymbol{\theta}}^k(\mathbf{y}_i^k) - \hat{h}_{\boldsymbol{\theta}}^k(\mathbf{x}_i^k). \quad (18)$$

Each feature is used exactly once for validation since the $\mathcal{D}_{\boldsymbol{\theta}}^k$ are disjoint. We make the simplifying assumption that splitting the original $n$ features into $K$ folds of $m$ features was possible without remainders. We can then order the $\mathbf{y}_i^k$ as

$$\mathbf{y}_1^1, \ldots, \mathbf{y}_m^1, \mathbf{y}_1^2, \ldots, \mathbf{y}_m^2, \mathbf{y}_1^3, \ldots, \mathbf{y}_m^K,$$

and relabel them from 1 to $n$. Doing the same for the $\mathbf{x}_i^k$, we obtain

$$J_n(\boldsymbol{\theta}) = \frac{1}{2} + \frac{1}{2n}\sum_{i=1}^n \hat{h}_{\boldsymbol{\theta}}^{k(i)}(\mathbf{y}_i) - \frac{1}{2n}\sum_{i=1}^n \hat{h}_{\boldsymbol{\theta}}^{k(i)}(\mathbf{x}_i). \quad (19)$$

The function $k(i)$ in the equation indicates to which validation set feature $i$ belonged. If the Bayes classification rule is used instead of the learned $\hat{h}_{\boldsymbol{\theta}}^{k(i)}$, we obtain $J_n^*(\boldsymbol{\theta})$ in Equation (3),

$$J_n^*(\boldsymbol{\theta}) = \frac{1}{2} + \frac{1}{2n}\sum_{i=1}^n h_{\boldsymbol{\theta}}^*(\mathbf{y}_i) - \frac{1}{2n}\sum_{i=1}^n h_{\boldsymbol{\theta}}^*(\mathbf{x}_i). \quad (20)$$

The function $k(i)$ disappeared because of the weak stationarity assumption that the marginal distributions of the $\mathbf{x}_i$ and $\mathbf{y}_i$ do not depend on $i$.

In what follows, it is helpful to introduce the set $H_{\boldsymbol{\theta}}^* = \{\mathbf{u} : h_{\boldsymbol{\theta}}^*(\mathbf{u}) = 1\}$. The normalized sums in (20) are then the fractions of features which belong to $H_{\boldsymbol{\theta}}^*$. Taking the expectation over $\mathbf{X}$ and $\mathbf{Y}_{\boldsymbol{\theta}}$, using that the expectation over the binary function $h_{\boldsymbol{\theta}}^*$ equals the probability of the set $H_{\boldsymbol{\theta}}^*$,

$$\mathrm{E}(h_{\boldsymbol{\theta}}^*(\mathbf{y}_i)) = \mathrm{P}_{\boldsymbol{\theta}}(H_{\boldsymbol{\theta}}^*), \qquad \mathrm{E}(h_{\boldsymbol{\theta}}^*(\mathbf{x}_i)) = \mathrm{P}_{\boldsymbol{\theta}^o}(H_{\boldsymbol{\theta}}^*), \quad (21)$$

we obtain the average discriminability $\mathrm{E}(J_n^*(\boldsymbol{\theta})) = J(\boldsymbol{\theta})$,

$$J(\boldsymbol{\theta}) = \frac{1}{2} + \frac{1}{2}\left(\mathrm{P}_{\boldsymbol{\theta}}(H_{\boldsymbol{\theta}}^*) - \mathrm{P}_{\boldsymbol{\theta}^o}(H_{\boldsymbol{\theta}}^*)\right). \quad (22)$$

The difference between $J_n$ and $J$ is twofold: First, relative frequencies instead of probabilities (expectations) occur. Second, learned classification rules instead of the Bayes classification rule are used.

*Remark:* There is an interesting analogy between the objective $J_n^*$ and the log-likelihood: The sum over the $\mathbf{y}_i$ does not depend on the observed data but on $\boldsymbol{\theta}$ and may be considered an analogue to the log-partition function (or an estimate of it). In the same analogy, the sum over the $\mathbf{x}_i$ corresponds to the logarithm of the unnormalized model of the data. The two terms have opposite signs and balance each other as in the methods for unnormalized models reviewed by Gutmann and Hyvärinen (2013a).

### A.2   Minimization of $J$

We note that $J(\boldsymbol{\theta}^o) = 1/2$. Since $H_{\boldsymbol{\theta}}^*$ contains only the points which are more probable under $P_{\boldsymbol{\theta}}$ than under $P_{\boldsymbol{\theta}^o}$, we have further that $J(\boldsymbol{\theta}) \geq 1/2$. Hence, $\boldsymbol{\theta}^o$ is a minimizer of $J$. However, $\boldsymbol{\theta}^o$ might not be the only one: Depending on the parametrization, it could be that $P_{\boldsymbol{\theta}^o} = P_{\boldsymbol{\theta}}$ for some $\tilde{\boldsymbol{\theta}}$ other than $\boldsymbol{\theta}^o$. We therefore made the identifiability assumption that the $\tilde{\boldsymbol{\theta}}$ are well separated from $\boldsymbol{\theta}^o$ so that there is is a compact subset $\Theta$ of the parameter space which contains $\boldsymbol{\theta}^o$ but none of the $\tilde{\boldsymbol{\theta}}$. The above can then be summarized as Proposition 2.

**PROPOSITION 2**  $J(\boldsymbol{\theta}^o) = 1/2$ *and* $J(\boldsymbol{\theta}) > 1/2$ *for all other* $\boldsymbol{\theta} \in \Theta$.

Restricting the parameter space to $\Theta$, consistency of $\hat{\boldsymbol{\theta}}_n$ follows from uniform convergence of $J_n$ to $J$ on $\Theta$ (van der Vaart 1998, Theorem 5.7).

### A.3   Uniform convergence of $J_n$ to $J$

We show that $J_n$ converges uniformly to $J$ if $J_n^*$ converges to $J$ and if $J_n$ stays close to $J_n^*$ for large $n$. This splits the convergence problem into two sub-problems with clear meanings which are discussed in the main text.

**PROPOSITION 3**

*If*   $\displaystyle\sup_{\boldsymbol{\theta} \in \Theta} |J(\boldsymbol{\theta}) - J_n^*(\boldsymbol{\theta})| \overset{P}{\to} 0$   *and*   $\displaystyle\sup_{\boldsymbol{\theta} \in \Theta} |J_n^*(\boldsymbol{\theta}) - J_n(\boldsymbol{\theta})| \overset{P}{\to} 0$

*then*   $\displaystyle\sup_{\boldsymbol{\theta} \in \Theta} |J(\boldsymbol{\theta}) - J_n(\boldsymbol{\theta})| \overset{P}{\to} 0.$  (23)

*Proof.*  By the triangle inequality, we have

$$|J(\boldsymbol{\theta}) - J_n(\boldsymbol{\theta})| \leq |J(\boldsymbol{\theta}) - J_n^*(\boldsymbol{\theta})| + |J_n^*(\boldsymbol{\theta}) - J_n(\boldsymbol{\theta})|, \quad (24)$$

so that

$$\sup_{\boldsymbol{\theta} \in \Theta} |J(\boldsymbol{\theta}) - J_n(\boldsymbol{\theta})| \leq \sup_{\boldsymbol{\theta} \in \Theta} |J(\boldsymbol{\theta}) - J_n^*(\boldsymbol{\theta})| + \sup_{\boldsymbol{\theta} \in \Theta} |J_n^*(\boldsymbol{\theta}) - J_n(\boldsymbol{\theta})|,$$

and hence

$$\mathrm{P}\left(\sup_{\boldsymbol{\theta} \in \Theta} |J(\boldsymbol{\theta}) - J_n(\boldsymbol{\theta})| > \epsilon\right) \leq \mathrm{P}\left(\sup_{\boldsymbol{\theta} \in \Theta} |J(\boldsymbol{\theta}) - J_n^*(\boldsymbol{\theta})| + \right.$$
$$\left. \sup_{\boldsymbol{\theta} \in \Theta} |J_n^*(\boldsymbol{\theta}) - J_n(\boldsymbol{\theta})| > \epsilon\right)  \quad (25)$$

It further holds that

$$\mathrm{P}\left(\sup_{\boldsymbol{\theta} \in \Theta} |J(\boldsymbol{\theta}) - J_n^*(\boldsymbol{\theta})| + \sup_{\boldsymbol{\theta} \in \Theta} |J_n^*(\boldsymbol{\theta}) - J_n(\boldsymbol{\theta})| > \epsilon\right) \leq \quad (26)$$
$$\mathrm{P}\left(\sup_{\boldsymbol{\theta} \in \Theta} |J(\boldsymbol{\theta}) - J_n^*(\boldsymbol{\theta})| > \frac{\epsilon}{2}\right) + \mathrm{P}\left(\sup_{\boldsymbol{\theta} \in \Theta} |J_n^*(\boldsymbol{\theta}) - J_n(\boldsymbol{\theta})| > \frac{\epsilon}{2}\right)$$

which concludes the proof.                                                                   $\square$

# Likelihood-free inference via classification
## – Supplementary material –

**Michael U. Gutmann**                    MICHAEL.GUTMANN@ED.AC.UK
*School of Informatics,*
*University of Edinburgh*

**Ritabrata Dutta**                        RITABRATA.DUTTA@USI.CH
*InterDisciplinary Institute of Data Science,*
*Universitá della Svizzera italiana*

**Samuel Kaski**                          SAMUEL.KASKI@AALTO.FI
*Helsinki Institute for Information Technology,*
*Department of Computer Science, Aalto University*

**Jukka Corander**                    JUKKA.CORANDERR@MEDISIN.UIO.NO
*Department of Biostatistics, University of Oslo*
*Helsinki Institute for Information Technology,*
*Department of Mathematics and Statistics, University of Helsinki*

## Contents

## List of Figures

## List of Tables

# 1. Models and algorithms

This section contains details on the classification methods, the models for continuous, binary, count and time-series data used to test our approach, as well as the ABC algorithm employed.

## 1.1 Classification methods

There are many possible classification methods, ranging from traditional logistic regression to more recent deep learning and kernel methods. For an introduction, we refer the reader to the textbooks by Wasserman (2004) and Hastie et al (2009). We used methods provided by two libraries: For linear and quadratic discriminant analysis (LDA and QDA), matlab's `classify.m` was employed. For $L_1$ and $L_2$ regularized polynomial logistic regression and support vector machine (SVM) classification, we used the `liblinear` classification library (Fan et al 2008), version 1.93, via the matlab interface, with a fixed regularization penalty (we used the default value $C = 1$). The `liblinear` library is for linear classification. Polynomial classification was implemented via polynomial basis expansion (Hastie et al 2009, Chapter 5). We rescaled the covariates to the interval $[-1, 1]$ and used the first nine Chebyshev polynomials of the first kind.

For all methods but LDA, multidimensional $\mathbf{x}_i$ were projected onto their principal components prior to classification and thereafter rescaled to variance one. This operation amounts to multiplying the $\mathbf{x}_i$ with a whitening matrix, and the $\mathbf{y}_i$ were multiplied with the same matrix.

The max-rule consisted in trying several classification methods and selecting the one giving the largest classification accuracy. We used $L_1$ and $L_2$ regularized polynomial logistic regression and SVM classification with the penalties $C = 0.1, 1, 10$, as well as LDA and QDA. When LDA was not applicable (as for the moving average model), it was excluded from the pool of classification methods used for the max-rule.

## 1.2 Models used for continuous, binary, count, and time series data

We tested the proposed inference method on several well-known distributions. This section details the models and lists the parameters used to generate the data, as well as the priors employed for Bayesian inference and the corresponding posterior distributions. The posterior distributions served as reference against which we compared the distributions produced by classifier ABC.

The sample average of $n$ data points $(x_1, \ldots, x_n)$ will be denoted by $\bar{x}$, and the sample variance by $s_n^2$,

$$\bar{x} = \frac{1}{n} \sum_{i=1}^{n} x_i, \qquad\qquad s_n^2 = \frac{1}{n} \sum_{i=1}^{n} (x_i - \bar{x})^2. \qquad\text{(S1)}$$

### 1.2.1 Continuous data

We considered inference for a univariate Gaussian with unknown mean and known variance, and inference of both mean and variance.

*Gaussian with unknown mean.* The data were sampled from a univariate Gaussian with mean $\mu^o = 1$ and variance $v^o = 1$. Inference was performed on the mean $\mu$. In the Bayesian setting, the prior distribution of $\mu$ was Gaussian,

$$\mu \sim \mathcal{N}(\mu_0, v_0), \qquad p(\mu|\mu_0, v_0) = \frac{1}{\sqrt{2\pi v_0}} \exp\left(-\frac{(\mu - \mu_0)^2}{2v_0}\right), \qquad \text{(S2)}$$

with mean $\mu_0 = 3$ and variance $v_0 = 1$. For Gaussian data with known variance $v^o$ and a Gaussian prior on the mean $\mu$, the posterior distribution of $\mu$ is Gaussian with mean $\mu_n$ and variance $v_n$,

$$\mu|\mathbf{X} \sim \mathcal{N}(\mu_n, v_n), \qquad \mu_n = \left(\frac{\mu_0}{v_0} + \frac{n\bar{x}}{v^o}\right)v_n, \qquad v_n = \left(\frac{1}{v_0} + \frac{n}{v^o}\right)^{-1}, \qquad \text{(S3)}$$

see, for example, (Gelman et al 2003, Chapter 2).

*Gaussian with unknown mean and variance.* The Gaussian data were generated with mean $\mu^o = 3$ and variance $v^o = 4$. Both mean $\mu$ and variance $v$ were considered unknown. In the Bayesian setting, the prior distribution was normal-inverse-gamma,

$$\mu|v \sim \mathcal{N}\left(\mu_0, \frac{v}{\lambda_0}\right), \quad v \sim \mathcal{G}^{-1}(\alpha_0, \beta_0), \quad p(v|\alpha_0, \beta_0) = \frac{\beta_0^{\alpha_0}}{\Gamma(\alpha_0)} v^{-\alpha_0 - 1} \exp\left(-\frac{\beta_0}{v}\right), \quad \text{(S4)}$$

where $\alpha_0$ and $\beta_0$ are the shape and scale parameters, respectively, and $\Gamma(.)$ is the gamma function, $\Gamma(t) = \int_0^\infty u^{t-1} \exp(-u) \mathrm{d}u$. The parameter values $\mu_0 = 0, \lambda_0 = 1, \alpha_0 = 3, \beta_0 = 0.5$ were used. This gives a prior variance with mean and standard deviation 0.25. The posterior is normal-inverse-gamma with updated parameters $\mu_n, \lambda_n, \alpha_n, \beta_n$,

$$\mu|v, \mathbf{X} \sim \mathcal{N}\left(\mu_n, \frac{v}{\lambda_n}\right), \quad \mu_n = \frac{\lambda_0 \mu_0 + n\bar{x}}{\lambda_0 + n}, \quad \lambda_n = \lambda_0 + n, \qquad \text{(S5)}$$

$$v|\mathbf{X} \sim \mathcal{G}^{-1}(\alpha_n, \beta_n), \quad \alpha_n = \alpha_0 + \frac{n}{2}, \qquad \beta_n = \beta_0 + \frac{n}{2}s_n^2 + \frac{n}{2}\frac{\lambda_0}{\lambda_0 + n}(\bar{x} - \mu_0)^2, \quad \text{(S6)}$$

see, for example, (Gelman et al 2003, Chapter 3).

### 1.2.2 BINARY DATA

The data were a random sample from a Bernoulli distribution with success probability (mean) $\mu^o = 0.2$. The prior on the mean $\mu$ was a beta distribution with parameters $\alpha_0 = \beta_0 = 2$,

$$\mu \sim \text{Beta}(\alpha_0, \beta_0), \qquad p(\mu|\alpha_0, \beta_0) = \frac{\Gamma(\alpha_0 + \beta_0)}{\Gamma(\alpha_0)\Gamma(\beta_0)} \mu^{\alpha_0 - 1}(1 - \mu)^{\beta_0 - 1}, \qquad \text{(S7)}$$

which has mean 0.5 and standard deviation 0.22. The posterior is beta with parameters $\alpha_n, \beta_n$,

$$\mu|\mathbf{X} \sim \text{Beta}(\alpha_n, \beta_n), \qquad \alpha_n = \alpha_0 + n\bar{x}, \qquad \beta_n = \beta_0 + n(1 - \bar{x}), \qquad \text{(S8)}$$

see, for example, (Gelman et al 2003, Chapter 2).

### 1.2.3 COUNT DATA

The data were a random sample from a Poisson distribution with mean $\lambda^o = 10$. The prior on the mean $\lambda$ was a gamma distribution with shape parameter $\alpha_0 = 3$ and rate parameter $\beta_0 = 1/2$,

$$\lambda \sim \mathcal{G}(\alpha_0, \beta_0), \qquad p(\lambda|\alpha_0, \beta_0) = \frac{\beta_0^{\alpha_0}}{\Gamma(\alpha_0)} \lambda^{\alpha_0 - 1} \exp\left(-\beta_0 \lambda\right). \tag{S9}$$

The prior distribution has mean 6, mode 4, and standard deviation 3.46. The posterior distribution is gamma with parameters $\alpha_n$, $\beta_n$,

$$\lambda|\mathbf{X} \sim \mathcal{G}(\alpha_n, \beta_n), \qquad \alpha_n = \alpha_0 + n\bar{x}, \qquad \beta_n = \beta_0 + n, \tag{S10}$$

see, for example, (Gelman et al 2003, Chapter 2).

### 1.2.4 TIME SERIES

We considered a moving average and an ARCH(1) model.

*Moving average model.* The time series is determined by the update equation

$$x_t = \epsilon_t + \theta \epsilon_{t-1}, \quad t = 1, \dots, T, \tag{S11}$$

where the $\epsilon_t, t = 0, \dots, T$ are independent standard normal random variables, and $\epsilon_0$ is unobserved. The observed data were generated with $\theta^o = 0.3$. The $\mathbf{x}_i$ for classification consisted of 2 consecutive time points $(x_t, x_{t+1})$.

For the derivation of the posterior distribution, it is helpful to write the update equation in matrix form. Let $\mathbf{x}_{0:T} = (x_0, \dots, x_T)$ and $\boldsymbol{\epsilon} = (\epsilon_0, \dots, \epsilon_T)$ be two column vectors of length $T+1$. The update equation does not specify the value of $x_0$. We thus set $x_0 = \epsilon_0$. Equation (S11) can then be written as

$$\mathbf{x}_{0:T} = \mathbf{B}\boldsymbol{\epsilon}, \qquad \mathbf{B} = \begin{pmatrix} 1 & 0 & & & \\ \theta & 1 & 0 & & \\ & \ddots & \ddots & \ddots & \\ & & \ddots & \ddots & 0 \\ & & & \theta & 1 \end{pmatrix}. \tag{S12}$$

It follows that $\mathbf{x}_{0:T}$ is zero mean Gaussian with covariance matrix $\mathbf{BB}^\top$. Since $\mathbf{x}_{0:T}$ has a Gaussian distribution, we can analytically integrate out the unobserved $x_0$. The resulting vector $\mathbf{x}_{1:T}$ is zero mean Gaussian with tridiagonal covariance matrix $\mathbf{C}$,

$$\mathbf{C} = \begin{pmatrix} 1+\theta^2 & \theta & & & \\ \theta & 1+\theta^2 & \theta & & \\ & \ddots & \ddots & \ddots & \\ & & \ddots & \ddots & \theta \\ & & & \theta & 1+\theta^2 \end{pmatrix}. \tag{S13}$$

We denote the distribution of $\mathbf{x}_{1:T}$ by $p(\mathbf{x}_{1:T}|\theta)$. A uniform prior on $(-1,1)$ was assumed for $\theta$. The posterior probability density function of $\theta$ given $\mathbf{x}_{1:T}$ is thus $p(\theta|\mathbf{x}_{1:T})$,

$$p(\theta|\mathbf{x}_{1:T}) = \frac{p(\mathbf{x}_{1:T}|\theta)}{\int_{-1}^{1} p(\mathbf{x}_{1:T}|\theta)\mathrm{d}\theta}, \qquad \theta \in (-1,1). \tag{S14}$$

The normalizing denominator can be computed using numerical integration. Numerical integration can also be used to compute the posterior mean and variance. We used matlab's `integral.m`.

*ARCH(1) model.* The model used was

$$x_t = \theta_1 x_{t-1} + \epsilon_t, \qquad \epsilon_t = \xi_t \sqrt{0.2 + \theta_2 \epsilon_{t-1}^2}, \qquad t = 1,\ldots,T, \qquad x_0 = 0, \tag{S15}$$

where the $\xi_t$ and $\epsilon_0$ are independent standard normal random variables. We call $\theta_1$ the mean process coefficient and $\theta_2$ the variance process coefficient. The observed data consist of the $x_t$ and we generated them with $(\theta_1^o, \theta_2^o) = (0.3, 0.7)$. The $\mathbf{x}_i$ used for classification consisted of 5 consecutive time points.

For the derivation of the posterior distribution, we introduce the column vectors $\boldsymbol{\epsilon} = (\epsilon_1, \ldots, \epsilon_T)$ and $\mathbf{x}_{1:T} = (x_1, \ldots, x_T)$ which are related by a linear transformation,

$$\boldsymbol{\epsilon} = \mathbf{Q}\mathbf{x}_{1:T}, \qquad\qquad \mathbf{Q} = \begin{pmatrix} 1 & 0 & & & \\ -\theta_1 & 1 & 0 & & \\ & \ddots & \ddots & \ddots & \\ & & \ddots & \ddots & 0 \\ & & & -\theta_1 & 1 \end{pmatrix}. \tag{S16}$$

Note that the band-diagonal matrix $\mathbf{Q}$ depends on $\theta_1$. The determinant of $\mathbf{Q}$ is one so that

$$p_{\mathbf{x}}(\mathbf{x}_{1:T}|\theta_1, \theta_2) = p_{\boldsymbol{\epsilon}}(\mathbf{Q}\mathbf{x}_{1:T}|\theta_1, \theta_2). \tag{S17}$$

The assumption on the $\xi_t$ implies that $\epsilon_t|\epsilon_{t-1}$ is Gaussian with variance $0.2 + \theta_2\epsilon_{t-1}^2$. We thus have

$$p_{\boldsymbol{\epsilon}}(\boldsymbol{\epsilon}|\theta_1, \theta_2) = p_1(\epsilon_1|\theta_1, \theta_2) \prod_{t=2}^{T} \frac{1}{\sqrt{2\pi(0.2 + \theta_2\epsilon_{t-1}^2)}} \exp\left(-\frac{\epsilon_t^2}{2(0.2 + \theta_2\epsilon_{t-1}^2)}\right), \tag{S18}$$

where $p_1$ is the pdf of $\epsilon_1$. Since $\epsilon_0$ is a latent variable following a standard normal distribution, $p_1$ is defined via an integral,

$$p_1(\epsilon_1|\theta_1, \theta_2) = \int \frac{1}{\sqrt{2\pi(0.2 + \theta_2\epsilon_0^2)}} \exp\left(-\frac{\epsilon_1^2}{2(0.2 + \theta_2\epsilon_0^2)}\right) \frac{1}{\sqrt{2\pi}} \exp\left(-\frac{\epsilon_0^2}{2}\right) \mathrm{d}\epsilon_0. \tag{S19}$$

We used numerical integration, matlab's `integral.m`, to evaluate it. The prior distribution of $(\theta_1, \theta_2)$ was the uniform distribution on the rectangle $(-1,1) \times (0,1)$. The posterior pdf $p(\theta_1, \theta_2|\mathbf{x}_{1:T})$ is

$$p(\theta_1, \theta_2|\mathbf{x}_{1:T}) = \frac{p_{\boldsymbol{\epsilon}}(\mathbf{Q}\mathbf{x}_{1:T}|\theta_1, \theta_2)}{\int_{-1}^{1} \int_{0}^{1} p_{\boldsymbol{\epsilon}}(\mathbf{Q}\mathbf{x}_{1:T}|\theta_1, \theta_2)\mathrm{d}\theta_1\mathrm{d}\theta_2}, \qquad (\theta_1, \theta_2) \in (-1,1) \times (0,1). \tag{S20}$$

The normalizing denominator, the posterior means and variances were computed with matlab's `integral2.m`.

### 1.3 ABC algorithm

There are several algorithms for approximate Bayesian computation (ABC, for an overview, see, for example, Marin et al 2012). For the results in the paper, we used a population Monte Carlo sampler, also known as sequential Monte Carlo ABC algorithm, with a Gaussian kernel (Marin et al 2012, Algorithm 4), (Beaumont et al 2009; Sisson et al 2007; Toni et al 2009). In brief, the algorithm starts with samples from the prior distribution and then produces sets (generations) of weighted independent samples where the samples from any given generation are the starting point to get the samples of the next generation. The empirical pdfs, scatter plots, and sample moments reported in the paper all take the weights into account.

In some ABC implementations, the acceptance thresholds are the empirical quantiles of the discrepancies of the accepted parameters; in others, a schedule is pre-defined. The pre-defined schedule depends on the scale of the discrepancy measure which is often unknown. Using quantiles avoids this problem, but if the quantile is set too low, too few samples will be accepted which results in a slow algorithm. For $J_n$, the scale is known. We took advantage of this and used a hybrid approach to choose the thresholds: The threshold for a generation was the maximum of the value given by a pre-defined schedule and the value given by the 0.1 quantile of the $J_n$ of the accepted parameters from the previous generation. With $t$ denoting the ABC generation, the schedule was $0.75/(1 + 0.45 \log t)$, which gives a value of 0.5 at $t = 3$.

Unlike a purely quantile-based approach, the hybrid approach avoids sudden jumps to small thresholds. We can thereby obtain posteriors for intermediate thresholds. These are faster to obtain but still informative. The final posteriors from both approaches are, however, very similar, as shown in Supplementary Figure 1.

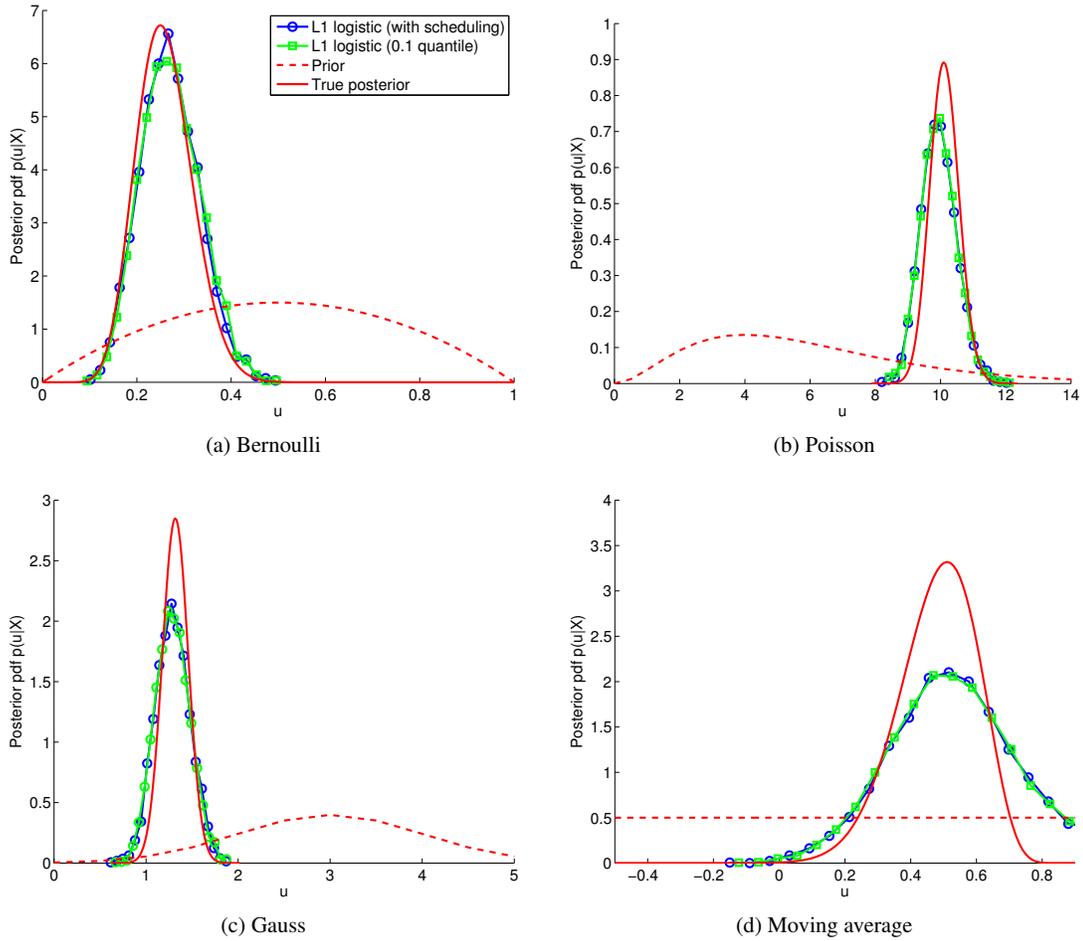

Supplementary Figure 1: Assessment of the hybrid approach to choose the acceptance thresholds in classifier ABC with a sequential Monte Carlo algorithm. The final posterior pdfs for the hybrid approach (blue, circles) and a purely quantile-based approach (green, squares) are very similar. The benefit of the hybrid approach is that it yields more quickly useful intermediate solutions. The results are for $L_1$-regularized polynomial logistic regression.

## 2. Measuring discrepancy via classification

In Figure 2 in the main text, chance-level discriminability was attained at a point close to the parameter $\boldsymbol{\theta}^o$ which was used to generate $\mathbf{X}$. We provide here two more such examples: Supplementary Figure 2 shows the results for a Gaussian distribution with unknown mean and variance, and Supplementary Figure 3 the results for the autoregressive conditional heteroskedasticity (ARCH) time series model in Equation (S15) with unknown mean and variance process coefficients. Parameter $\boldsymbol{\theta}^o$ is marked with a red cross.

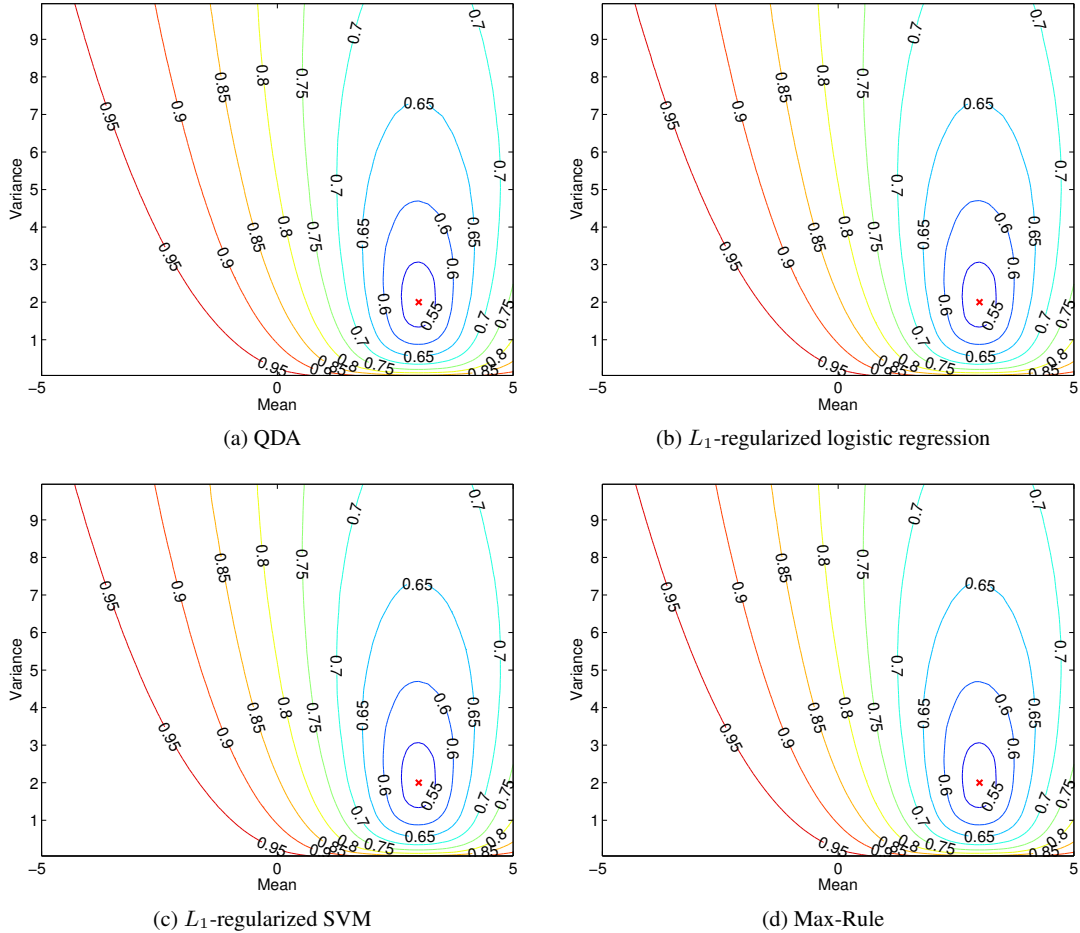

Supplementary Figure 2: Gaussian with unknown mean and variance. The contour plots show $J_n$ as a function of the two parameters for large sample sizes ($n = 100,000$). The different panels depict results for different classification methods. All obtain their minimal classification accuracy, chance-level discriminability 0.5, close to $\boldsymbol{\theta}^o$.

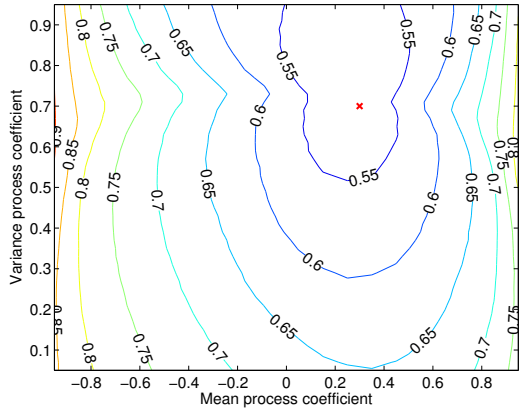

(a) QDA

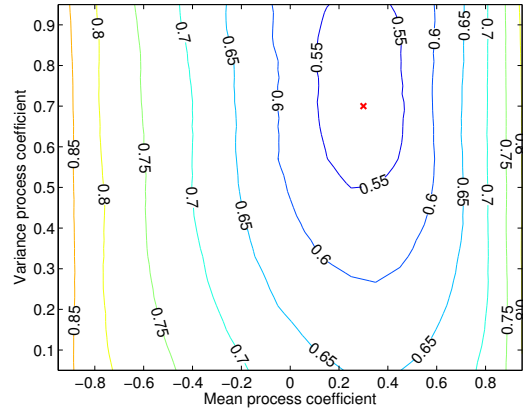

(b) $L_1$-regularized logistic regression

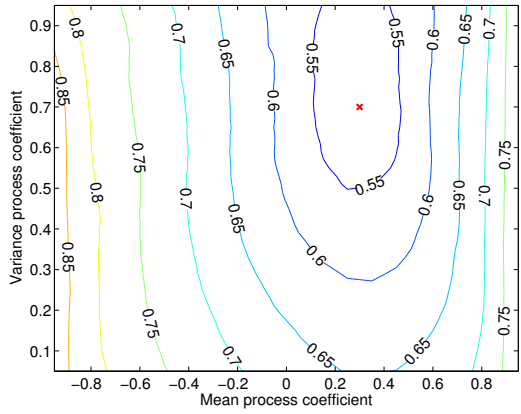

(c) $L_1$-regularized SVM

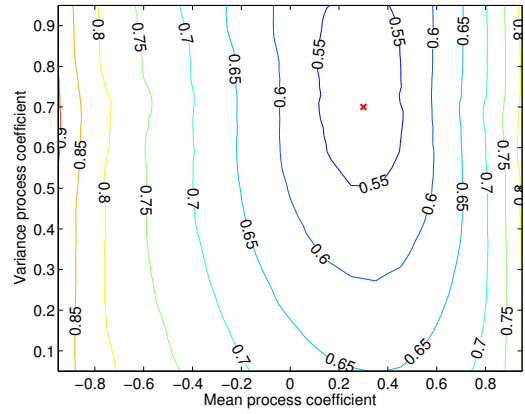

(d) Max-Rule

Supplementary Figure 3: ARCH(1) model in Equation (S15) with unknown mean and variance process coefficients $\theta_1$ and $\theta_2$. The results are for $n = 10{,}000$ and visualized as in Supplementary Figure 2.

## 3. Classical inference via classification

In Figure 3 in the main text, we plotted the mean squared estimation error $\mathrm{E}[||\hat{\boldsymbol{\theta}}_n - \boldsymbol{\theta}^o||^2]$ for the examples in Figure 2 against the sample size $n$ for $L_1$-regularized logistic regression. Supplementary Figure 4 shows the corresponding results for linear discriminant analysis (LDA), quadratic discriminant analysis (QDA), $L_1$-regularized polynomial support vector machine (SVM) classification, and the max-rule. As for the results in the main text, the decay is linear on the log-log scale which suggests convergence in quadratic mean, hence convergence in probability, and thus consistency of $\hat{\boldsymbol{\theta}}_n$.

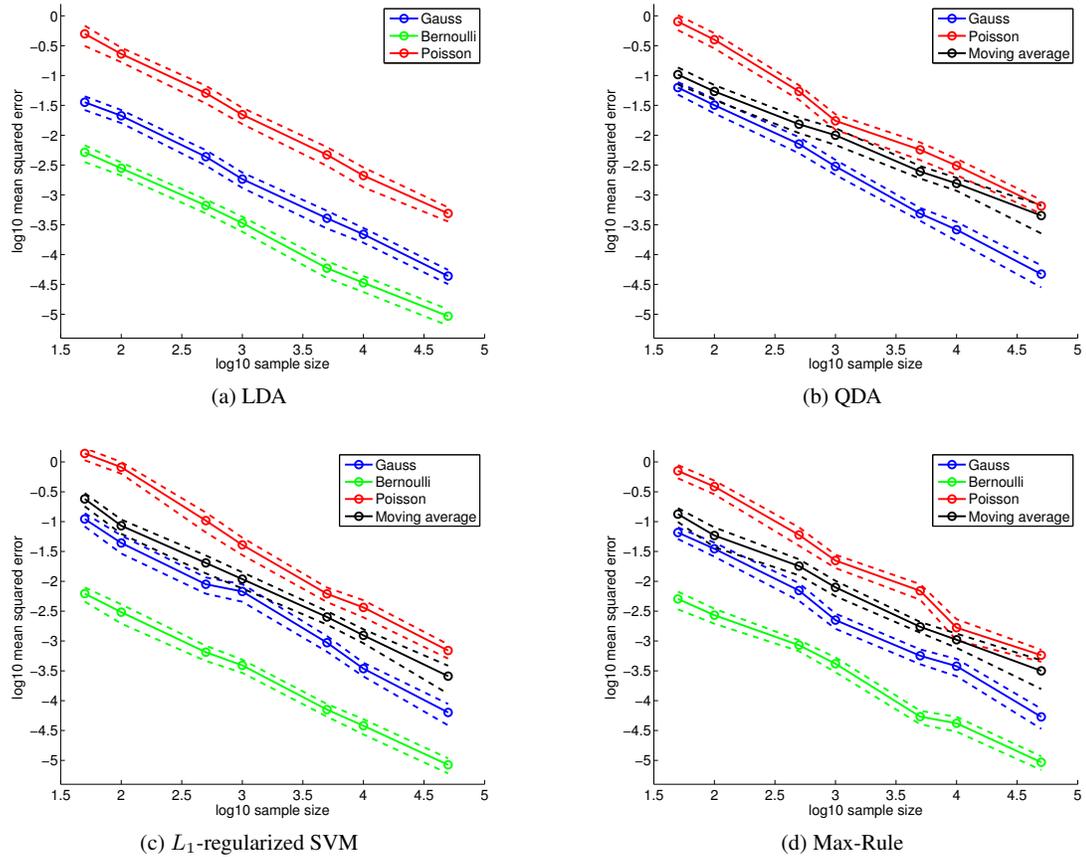

Supplementary Figure 4: The mean squared estimation error for the examples in Figure 2 in the main text as a function of the sample size $n$ (solid lines, circles). The mean was computed as an average over 100 outcomes. The dashed lines depict the mean $\pm$ 2 standard errors. For QDA, the Bernoulli case is not reported because, sometimes, data with degenerate covariance matrices were generated, which the standard QDA algorithm used was not able to handle. For LDA, the moving average case was omitted since LDA cannot approximate its Bayes classification rule as discussed in the main text. The linear trend on the log-log scale suggests convergence in quadratic mean, and hence consistency of the estimator $\hat{\boldsymbol{\theta}}_n$.

## 4. Bayesian inference via classification

This section contains further results for classifier ABC on data with known properties, supplementing Section 4 of the main text.

### 4.1 The inferred posterior distributions for all classification methods used

We report the posterior distributions for all classification methods used in the paper in Supplementary Figure 5 to Supplementary Figure 10. The results are organized according to the modality of the data.

The results are for $n = 50$ and 10,000 ABC samples with a sequential Monte Carlo implementation of ABC. For the univariate cases, empirical pdfs of the ABC samples are shown together with the reference posterior pdf (red solid) and the prior pdf used (red dashed). For the bivariate cases, the ABC samples are shown as a scatter plot and the reference posterior is visualized using contour plots (red solid line). The priors are either shown as contour plots (with red dashed lines) or, if uniform, by hatching their domain.

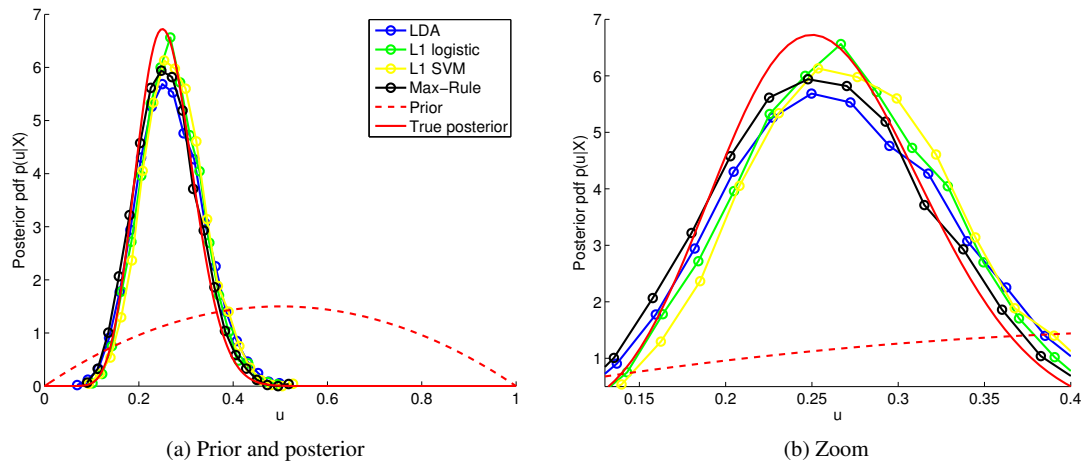

(a) Prior and posterior        (b) Zoom

Supplementary Figure 5: Binary data: Inferred posterior distribution of the success probability of a Bernoulli random variable.

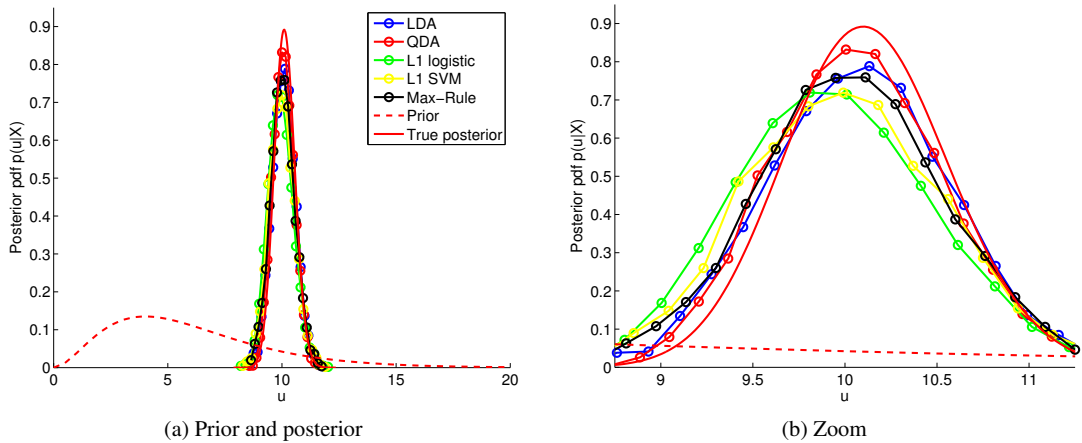

(a) Prior and posterior        (b) Zoom

Supplementary Figure 6: Count data: Inferred posterior distribution of the mean of a Poisson random variable.

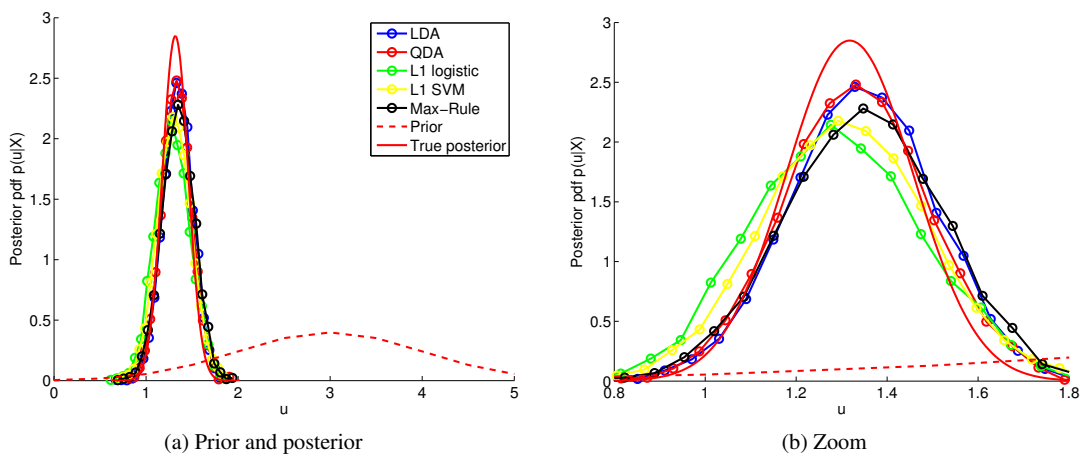

(a) Prior and posterior        (b) Zoom

Supplementary Figure 7: Continuous data: Inferred posterior distribution of the mean of a Gaussian random variable with known variance.

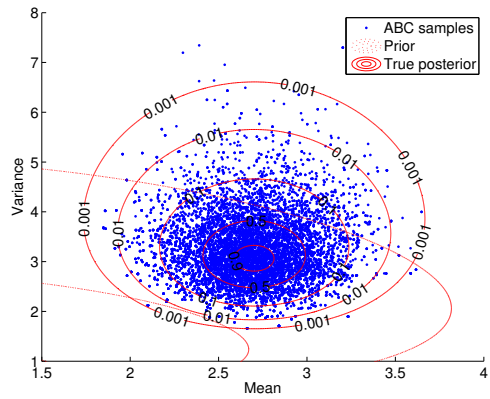

(a) QDA

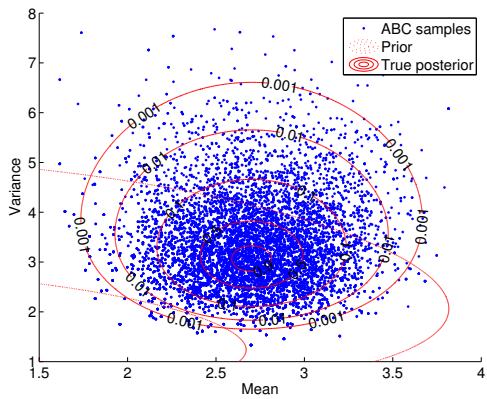

(b) $L_1$-regularized logistic regression

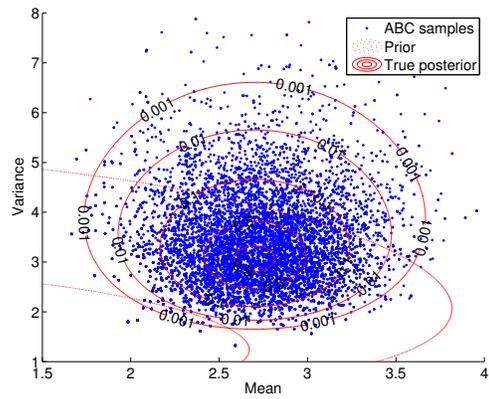

(c) $L_1$-regularized SVM

Supplementary Figure 8: Continuous data: Inferred posterior distribution of the mean and variance of a Gaussian random variable.

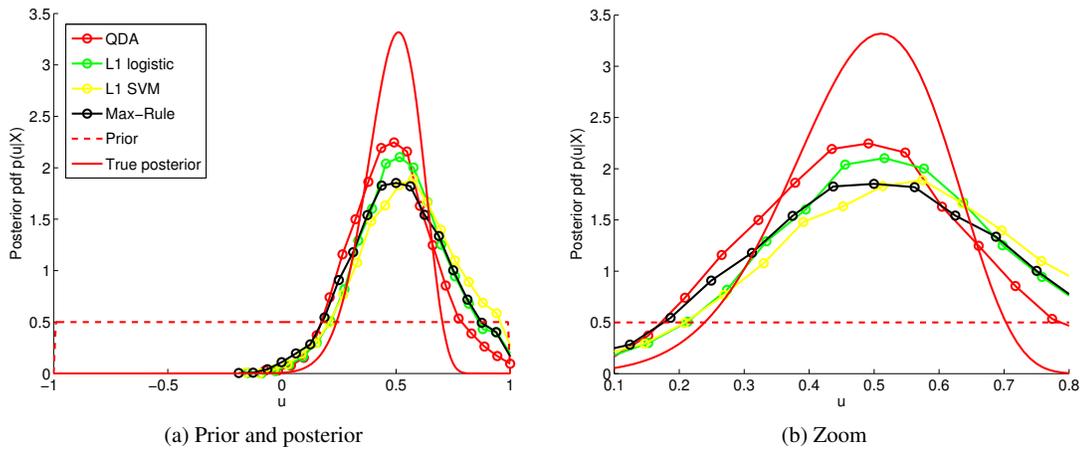

(a) Prior and posterior

(b) Zoom

Supplementary Figure 9: Time series: Inferred posterior distribution of the lag coefficient of a zero mean moving average model of order one.

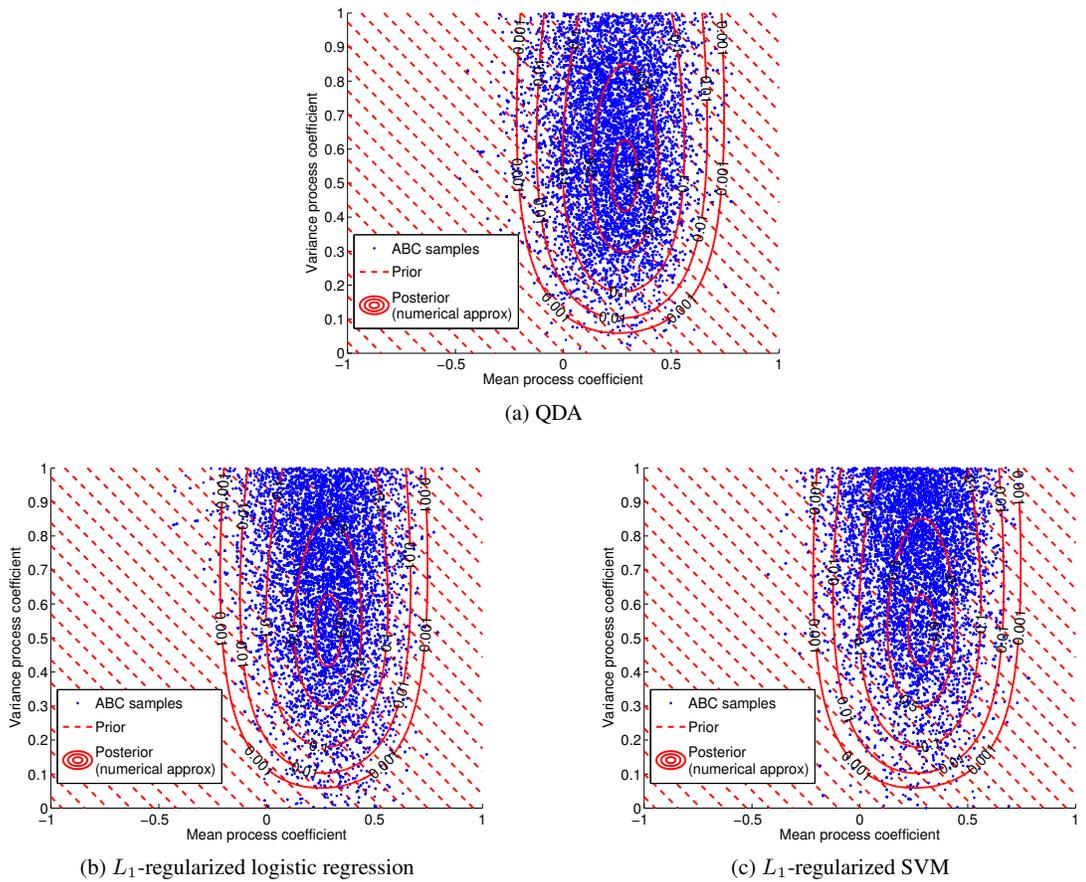

(a) QDA

(b) $L_1$-regularized logistic regression

(c) $L_1$-regularized SVM

Supplementary Figure 10: Time series: Inferred posterior distribution of the mean and variance process coefficients of a ARCH(1) model.

## 4.2 Movies showing the evolution of the inferred posteriors

The sequential Monte Carlo algorithm which we used together with classifier ABC is iteratively morphing a prior distribution into a posterior distribution. Table 1 contains links to movies which show this process.

| Data | LDA | QDA | Logi regr | SVM | Max-Rule |
|---|---|---|---|---|---|
| Binary (Bernoulli) | avi mp4 | | avi mp4 | avi mp4 | avi mp4 |
| Count (Poisson) | avi mp4 | avi mp4 | avi mp4 | avi mp4 | avi mp4 |
| Continuous (Gauss, mean) | avi mp4 | avi mp4 | avi mp4 | avi mp4 | avi mp4 |
| Continuous (Gauss, mean & var) | | avi mp4 | avi mp4 | avi mp4 | avi mp4 |
| Time series (moving average) | | avi mp4 | avi mp4 | avi mp4 | avi mp4 |
| Time series (ARCH) | | avi mp4 | avi mp4 | avi mp4 | avi mp4 |

Table 1: Links to movies showing the inference process of classifier ABC with a sequential Monte Carlo algorithm. Available online from the homepage of the first author.

## 4.3 Relative errors in posterior means and standard deviations

As a quantitative analysis, we computed the relative error in the mean and the standard deviation of the inferred posterior distributions. The comparison is based on the mean and standard deviation of the true posterior if available, or, if not, the posterior obtained by deterministic numerical integration, see Supplementary material 1.2.

Supplementary Figure 11 shows the relative error for the max-rule as a function of the iteration in the ABC algorithm. The error stabilizes within 4-5 iterations. For the examples with independent data points, the errors in the posterior mean are within 5% after stabilization. A larger error of 15% occurs for the time series data. The histograms and scatter plots show, however, that the corresponding ABC samples are still very reasonable.

While the relative error for the mean is both positive or negative, for the standard deviation, the error is positive only. This means that the inferred posteriors have a larger spread than the reference posteriors, that is, the posterior variance is overestimated. Further, the relative errors are generally larger for the standard deviations than for the means. This may not be too surprising though: Also in the framework of maximum likelihood estimation, the variance of the estimate of the variance is twice the variance of the estimate of the mean for standard normal random variables.

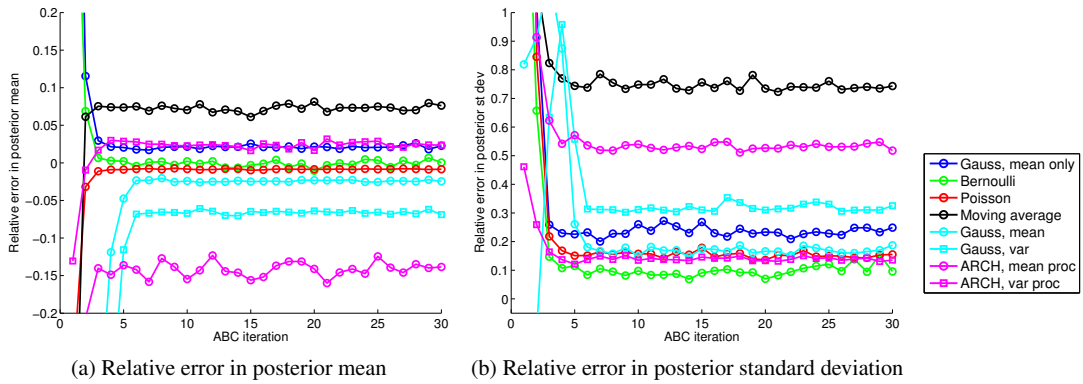

(a) Relative error in posterior mean

(b) Relative error in posterior standard deviation

Supplementary Figure 11: Quantitative analysis of the inferred posterior distributions. The curves show the relative error in the posterior mean and standard deviation for the Gauss, Bernoulli, Poisson, moving average, and ARCH examples. The results are for classification with the max-rule.

# 5. Application on real data

This section supplements Section 5 of the main text. Further results and analysis of our application to infectious disease epidemiology are presented.

## 5.1 Evolution of inferred posterior distributions on simulated data

We inferred the individual-based epidemic model with a sequential Monte Carlo ABC with $J_n$ as discrepancy measure (classifier ABC). Supplementary Figure 12 visualizes the evolution of the inferred posterior distribution over four generations. We show the results for classifier ABC with random subsets (blue, circles) and without (red, squares). For reference, the results with the method by Numminen et al (2013), which uses expert knowledge, are shown in black (point markers). Figure 7 in the main text shows the fourth generation results in greater detail.

Numminen et al (2013) presented posterior distributions for four generations. In both the results reported here and the results by Numminen et al (2013), the mean of the inferred posteriors seems to have stabilized after four generations. The spread of the inferred posteriors, however, is still slightly shrinking. We thus ran the simulations for an additional fifth iteration. The results are shown in Supplementary Figure 13. With the fifth iteration, the posterior pdfs for classifier ABC with random projections became more concentrated and also more similar to the expert solution than the posteriors of classifier ABC without random projections. The smaller posterior variance is in line with the tighter $J_n$-diagrams in Figure 6 in the main text.

## 5.2 Evolution of inferred posterior distributions on real data

The evolution of the posterior pdfs during the ABC algorithm is shown in Supplementary Figure 14. Starting from uniform distributions, posterior distributions with well defined modes emerged. Figure 8 in the main text shows the fourth generation results in greater detail. While the posteriors of $\Lambda$ and $\theta$ are qualitatively similar for all three methods, the posterior of $\beta$ has a smaller mode for classifier ABC with random subsets (blue, crosses) than for classifier ABC without random subsets (red, asterisks) or the expert solution (black, plus markers). This behavior persists in the fifth generation as shown in Supplementary Figure 15. Compared to the fourth generation results, the posteriors for classifier ABC with random subsets (blue, crosses) and the expert solution (black, plus markers) became in the fifth generation more concentrated than the posterior for classifier ABC without random subsets (red, asterisks).

## 5.3 Further results on compensating missing expert statistics with classifier ABC

Classifier ABC, or more generally the discrepancy measure $J_n$, is able to incorporate expert statistics, by letting them be features (covariates) in the classification. On the one hand, this allows for expert knowledge to be used in classifier ABC. On the other hand, it allows one to enhance expert statistics by data-driven choices. The latter is particularly important if only a insufficient set of summary statistics may be specified. We show here that classifier ABC can counteract shortcomings caused by a suboptimal choice of expert statistics, thereby making the inference more robust.

We selected two (simple) expert statistics used by Numminen et al (2013), namely the number of different strains circulating and the proportion of individuals who are infected. We then inferred the posteriors with this reduced set of summary statistics only, using the method of Numminen et al (2013). Supplementary Figure 16 visualizes the resulting posterior pdfs (curves in magenta with diamond markers). A comparison with the expert solution with a full set of summary statistics (black curve, point markers) shows that the posterior distributions of $\Lambda$ and $\theta$ are affected by the suboptimal choice of expert statistics. We then included the two selected expert statistics as additional features in classifier ABC. Consequently, the posteriors of $\Lambda$ and $\theta$ recuperated, both when random features were present (cyan curve with triangles) or not (red curve with hexagrams).

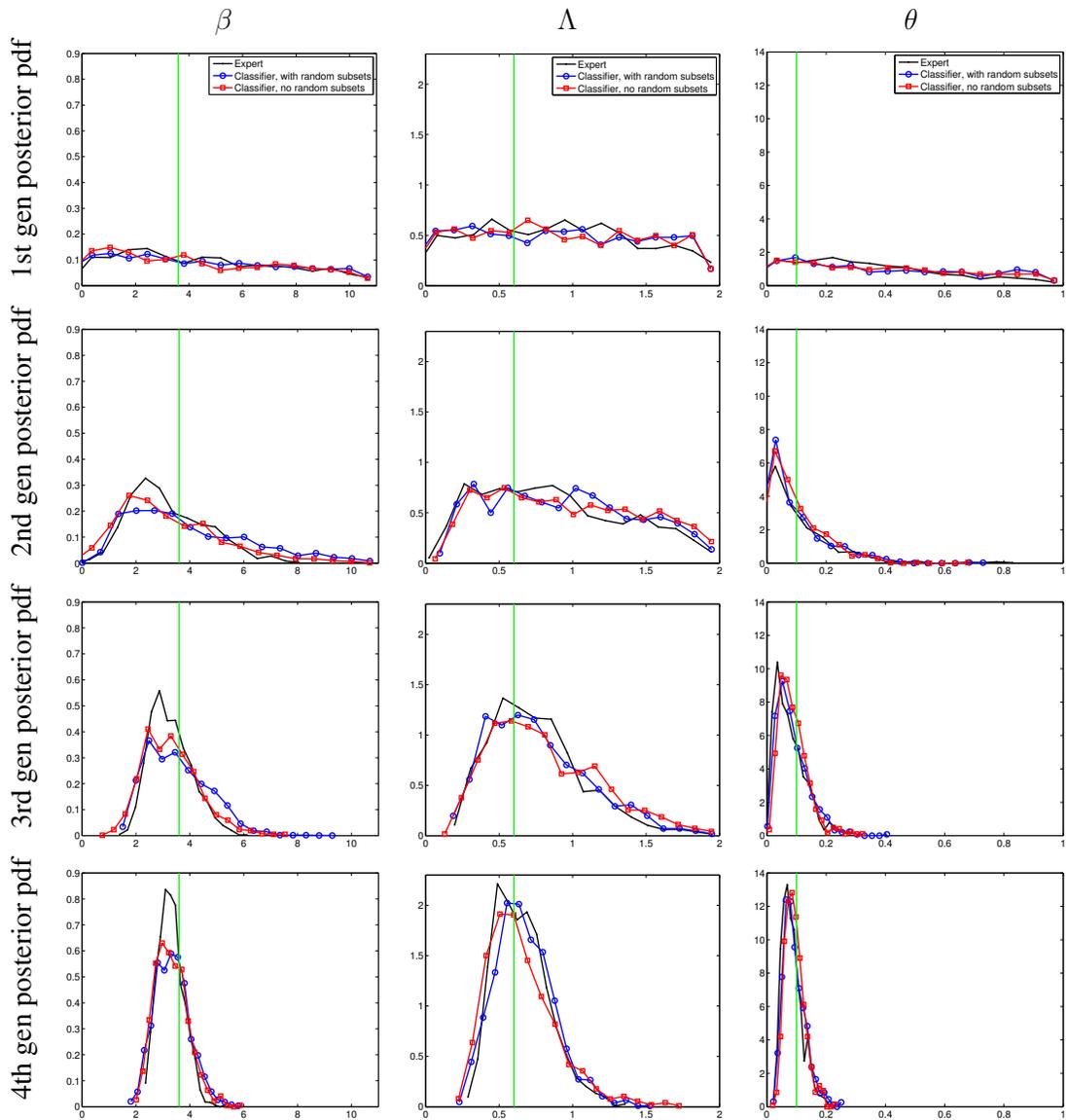

Supplementary Figure 12: Simulated data: Evolution of the posterior pdfs (scaled histograms of the samples). Black, points: ABC solution using expert knowledge, produced with code from Numminen et al (2013). Blue, circles: classifier ABC with random subsets. Red, squares: classifier ABC without random subsets. Green vertical lines: location of the data generating parameter $\theta^o$. The results are for 1,000 ABC samples.

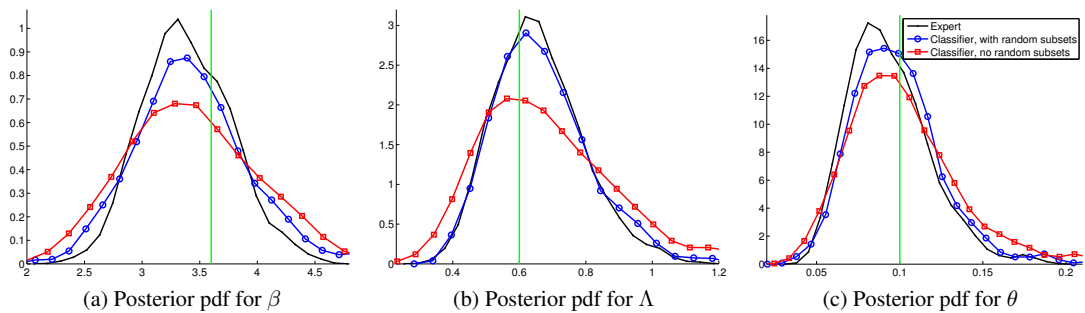

(a) Posterior pdf for $\beta$  (b) Posterior pdf for $\Lambda$  (c) Posterior pdf for $\theta$

Supplementary Figure 13: Simulated data: Fifth generation results. The posterior pdfs are kernel density estimates based on 1,000 ABC samples. We used matlab's ksdensity.m with the default settings, that is, a Gaussian kernel with an adaptively chosen bandwidth. Classifier ABC with random projections (blue, circles) yielded results which are more similar to the expert solution (black, points) than classifier ABC without random projections (red, squares).

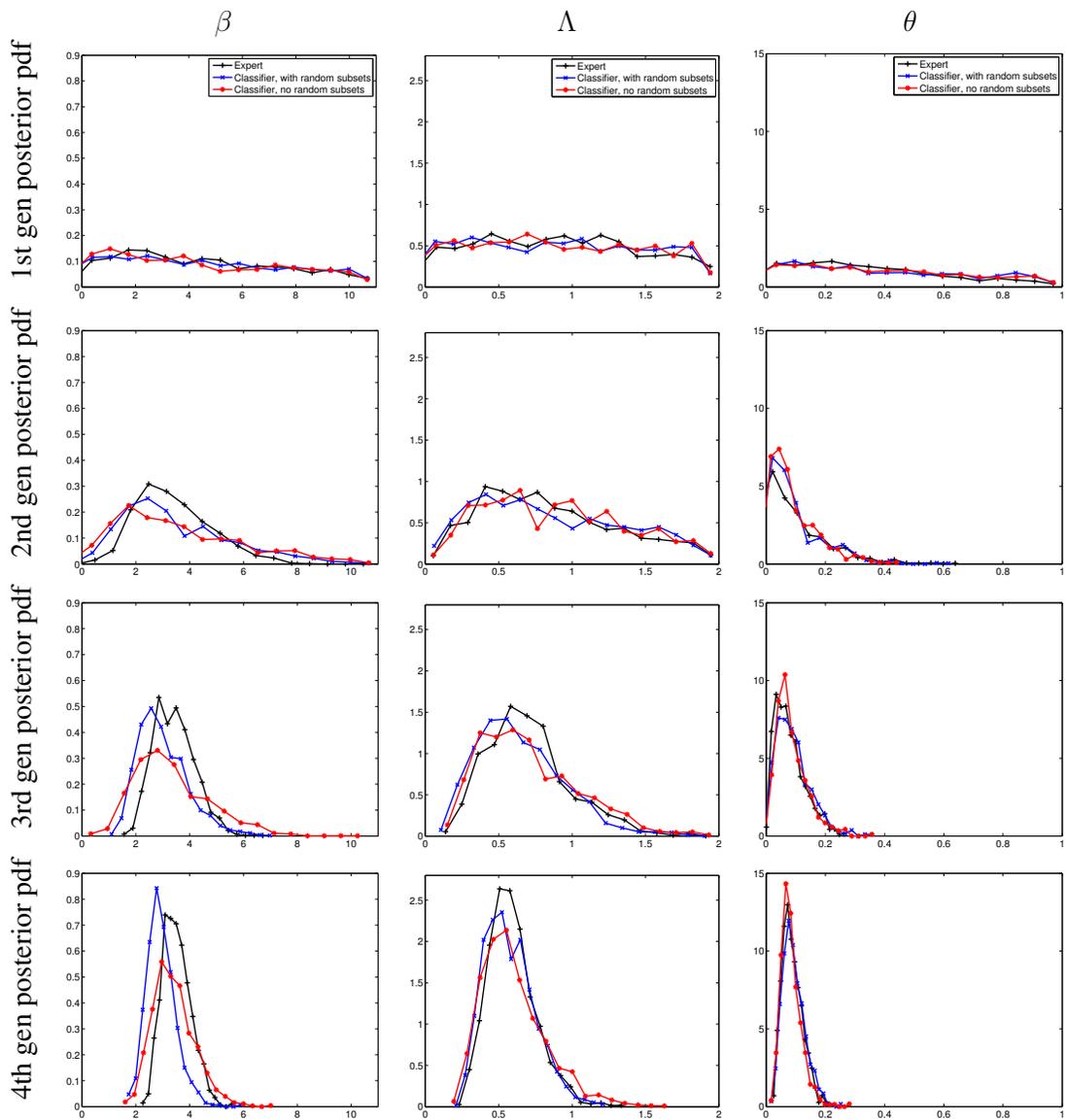

Supplementary Figure 14: Real data: Evolution of the posterior pdfs (scaled histograms of the samples). Black, plus markers: ABC solution using expert knowledge, produced with code from Numminen et al (2013). Blue, crosses: classifier ABC with random subsets. Red, asterisks: classifier ABC without random subsets. The results are for 1,000 ABC samples.

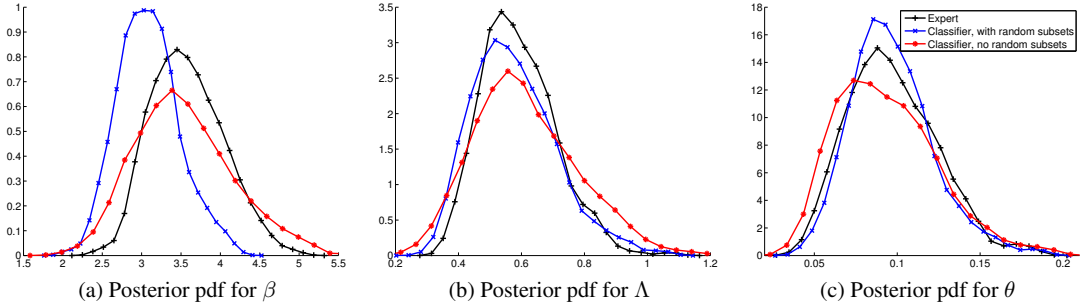

(a) Posterior pdf for $\beta$     (b) Posterior pdf for $\Lambda$     (c) Posterior pdf for $\theta$

Supplementary Figure 15: Real data: Fifth generation results. The posterior pdfs are kernel density estimates based on 1,000 ABC samples. We used matlab's ksdensity.m with the default settings, that is, a Gaussian kernel with an adaptively chosen bandwidth. The posteriors for classifier ABC with random subsets (blue, crosses) and the expert solution (black, plus markers) are more concentrated than the posterior for classifier ABC without random subsets (red, asterisks).

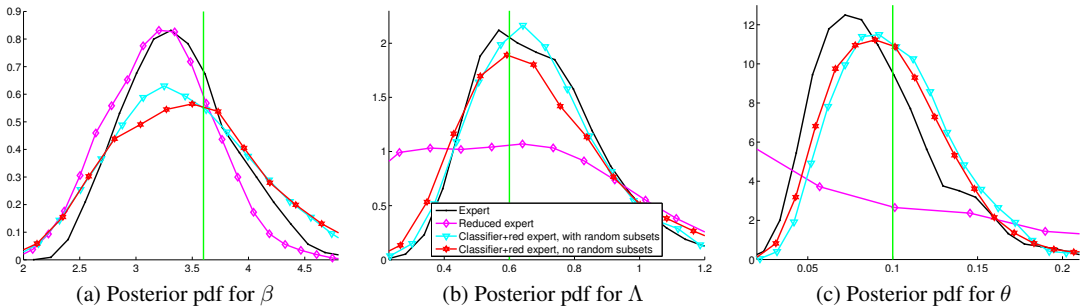

(a) Posterior pdf for $\beta$     (b) Posterior pdf for $\Lambda$     (c) Posterior pdf for $\theta$

Supplementary Figure 16: Using expert statistics in classifier ABC. The results are for simulated data and show the fourth generation pdfs. Visualization is as in e.g. Supplementary Figure 15. ABC with a reduced set of expert statistics affected the posteriors (black curve with points vs magenta curve with diamonds as markers). Classifier ABC was able to counteract the shortcomings caused by the suboptimal choice of expert statistics (cyan curve with triangles and red curve with hexagrams).

23